\documentclass[%
superscriptaddress,
showpacs,preprintnumbers,
 amsmath,amssymb,
 aps,twocolumn,
sort&compress,nofootinbib]{revtex4-1}

\usepackage[utf8]{inputenc}  
\usepackage[T1]{fontenc}

\usepackage{comment}
\usepackage{graphicx}
\usepackage{dcolumn}
\usepackage{bm}
\usepackage{enumitem}
\usepackage{color}
\usepackage{multirow}
\usepackage{adjustbox}	
\usepackage{placeins}	

\newcommand{\fig}[1]{Figure~\ref{#1}}
\newcommand{\tab}[1]{Table~\ref{#1}}
\newcommand{\Sec}[1]{Section~\ref{#1}}
\newcommand{\eq}[1]{Eq.~(\ref{#1})}
\newcommand{\Ref}[1]{Ref.~\cite{#1}}
\newcommand{\Alpg}{$^{27}\mathrm{Al(p,}\gamma)^{28}\mathrm{Si}$ }

\graphicspath{{../Figures/}}

\newcommand{\hzdr}{\affiliation{Helmholtz-Zentrum Dresden-Rossendorf (HZDR), 01328 Dresden, Germany}}
\newcommand{\tu}{\affiliation{Technische Universität Dresden, 01069 Dresden, Germany}}
\newcommand{\pd}{\affiliation{Dipartimento di Fisica e Astronomia, Universit\`a degli studi di Padova, 35131 Padova, Italy}}
\newcommand{\infn}{\affiliation{Istituto Nazionale di Fisica Nucleare (INFN), Sezione di Padova, 35131 Padova, Italy}}

\newcommand{\lulea}{\affiliation{Department of Engineering Sciences and Mathematics, Lule{\aa} University of Technology, 97187 Lule{\aa}, Sweden}}

\begin{document}


\title{Astrophysical S-factor of the $^{14}\textrm{N(p,}\gamma\textrm{)}^{15}\textrm{O}$ reaction at 0.4 -- 1.3\,MeV}

\author{L.~Wagner}%
\hzdr
\tu

\author{S.~Akhmadaliev}%
\hzdr

\author{M.~Anders}%
\hzdr
\tu

\author{D.~Bemmerer}%
\email{d.bemmerer@hzdr.de}%
\hzdr

\author{A.~Caciolli}%
\pd
\infn

\author{St.~Gohl}%
\hzdr
\tu

\author{M.~Grieger}%
\hzdr
\tu

\author{A.~Junghans}%
\hzdr

\author{M.~Marta}%
\affiliation{GSI Helmholtzzentrum für Schwerionenforschung, D-64291 Darmstadt, Germany}

\author{F.~Munnik}%
\hzdr

\author{T.~P.~Reinhardt}%
\tu

\author{S.~Reinicke}%
\hzdr
\tu


\author{M.~Röder}%
\hzdr
\tu

\author{K.~Schmidt\thanks{Present address: National Superconducting Cyclotron Laboratory, Michigan State University, East Lansing, Michigan 48824, USA}}%
\hzdr
\tu

\author{R.~Schwengner}%
\hzdr

\author{M.~Serfling}%
\hzdr
\tu

\author{M.~P.~Tak\'acs}%
\hzdr
\tu

\author{T.~Szücs}%
\hzdr

\author{A.~Vomiero}%
\lulea

\author{A.~Wagner}%
\hzdr

\author{K.~Zuber}%
\tu

\date{\today}

\begin{abstract}
The $^{14}\textrm{N(p,}\gamma\textrm{)}^{15}\textrm{O}$ reaction is the slowest reaction of the carbon-nitrogen cycle of hydrogen burning and thus determines its rate. The precise knowledge of its rate is required to correctly model hydrogen burning in asymptotic giant branch stars. In addition, it is a necessary ingredient for a possible solution of the solar abundance problem by using the solar $^{13}$N and $^{15}$O neutrino fluxes as probes of the carbon and nitrogen abundances in the solar core. After the downward revision of its cross section due to a much lower contribution by one particular transition, capture to the ground state in $^{15}$O, the evaluated total uncertainty is still 8\%, in part due to an unsatisfactory knowledge of the excitation function over a wide energy range. The present work reports precise S-factor data at twelve energies between 0.357-1.292~MeV for the strongest transition, capture to the 6.79~MeV excited state in $^{15}$O, and at ten energies between 0.479-1.202~MeV for the second strongest transition, capture to the ground state in $^{15}$O. An R-matrix fit is performed to estimate the impact of the new data on astrophysical energies. The recently suggested slight enhancement of the 6.79~MeV transition at low energy could not be confirmed. The present extrapolated zero-energy S-factors are $S_{6.79}(0)$~=~1.24$\pm$0.11~keV~barn and $S_{\rm GS}(0)$~=~0.19$\pm$0.05~keV~barn.

\end{abstract}


\maketitle

\section{\label{sec:intro}Introduction}

The rate of the carbon--nitrogen--oxygen (CNO) cycle of hydrogen burning plays a crucial role in stellar models, both for energy generation and for nucleosynthetic predictions \cite{Wiescher10-ARNPS}. Once the cycle has reached equilibrium, its rate is determined by the rate of the slowest reaction, $^{14}\textrm{N(p,}\gamma\textrm{)}^{15}\textrm{O}$. 

\begin{figure}[b]
\includegraphics[width=\columnwidth]{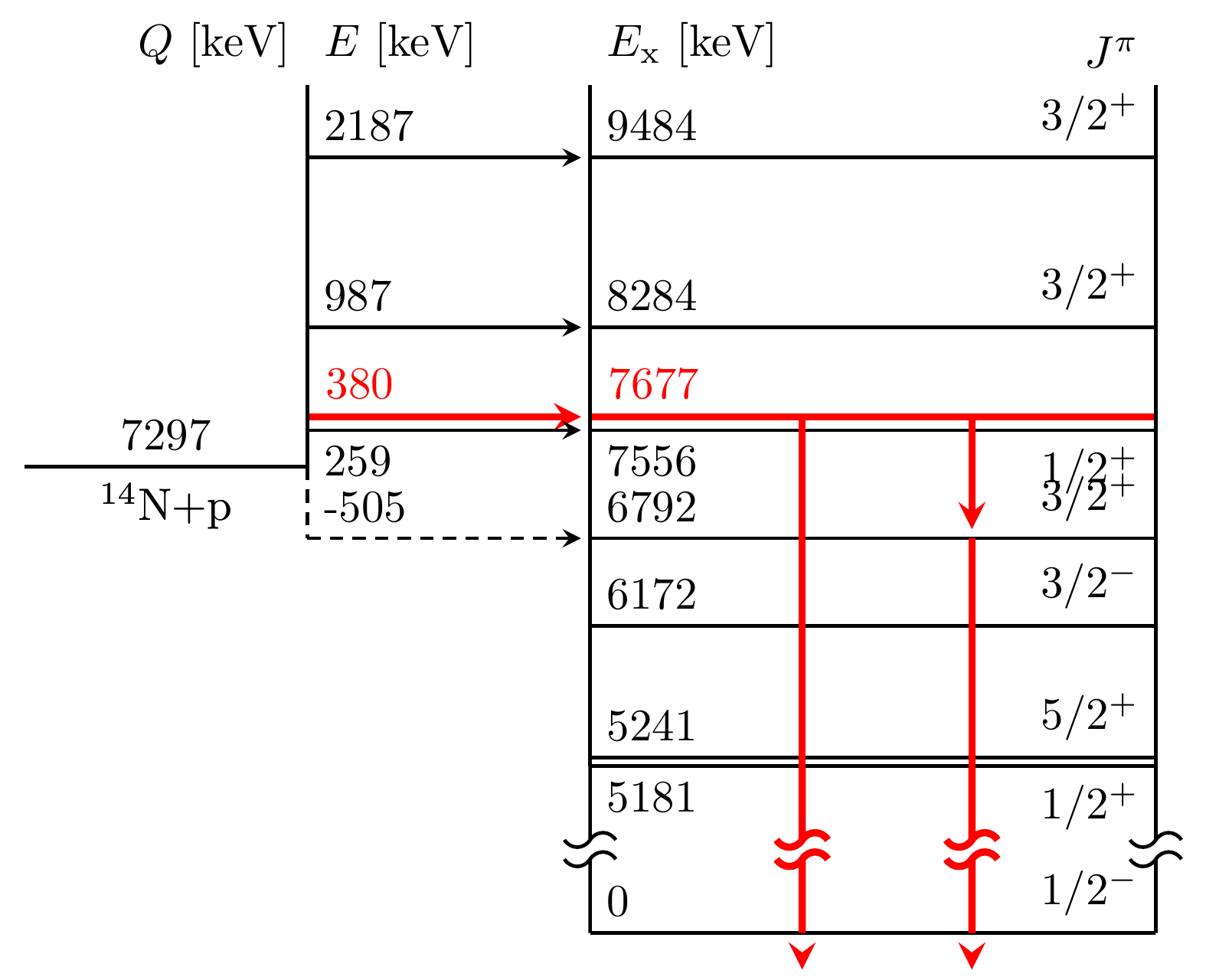}
	\caption{\label{fig:termschema} Level scheme of $^{15}$O \cite{Ajzenberg13_15_91-NPA,Imbriani05-EPJA}. The strongest transitions in the $^{14}\textrm{N(p,}\gamma\textrm{)}^{15}\textrm{O}$ reaction at 0.4-1.4\,MeV proton beam energies are marked with red arrows, using a center-of-mass-energy of 380\,keV as an example.}
\end{figure}

The $^{14}\textrm{N(p,}\gamma\textrm{)}^{15}\textrm{O}$ reaction proceeds by capture to the ground state and several excited states in the $^{15}$O nucleus (Fig.~\ref{fig:termschema}). Its cross section $\sigma(E)$ can be parameterized as the astrophysical S-factor $S(E)$ \cite{Iliadis15-Book}, which is given by the relation:
\begin{equation} 
S(E) = \sigma(E) E \exp\left[\frac{212.4}{\sqrt{E \, [{\rm keV}]}} \right]
\end{equation} 
where $E$ is the center-of-mass energy.

Both the central value and the uncertainty of the $^{14}\textrm{N(p,}\gamma\textrm{)}^{15}\textrm{O}$ reaction rate are of significance for a number of astrophysical scenarios, such as hydrogen shell burning in asymptotic giant branch stars \cite{Herwig04-ApJL,Herwig06-PRC}, the dating of globular clusters \cite{Imbriani04-AA,Innocenti04-PLB}, and the solar abundance problem \cite{Haxton08-ApJ}.

The latter problem has arisen due to the re-determination of the elemental abundances in the sun based on three-dimensional models for the solar atmosphere, which entailed a significant reduction of the adopted abundance values \cite{Asplund09-ARAA,Caffau11-SolPhys}. When fed into the standard solar model, the new, lower abundances lead to a predicted sound speed profile that is at odds with helioseismological observations \cite{Serenelli09-ApJL,Basu15-SSRv}. This conflict between two observables, i.e. elemental abundances and helioseismology, may in principle be addressed by studying an independent third observable. 

It has been suggested \cite{Haxton08-ApJ} to use solar neutrinos from the $\beta^+$ decay of the CNO cycle nuclides $^{13}$N, $^{15}$O, and $^{17}$F for this purpose. These neutrinos may in principle be detected at modern neutrino detectors like Borexino \cite{Borexino14-Nature}, SNO+ \cite{Andringa16-AHEP}, and  possibly at the Chinese Jinping Underground Facility \cite{Beacom17-CPC}. Using the well-measured $^8$B neutrino flux as a solar thermometer \cite{Takacs15-PRD}, the CNO neutrino flux would be directly proportional to the abundances of carbon, nitrogen, and oxygen in the solar core \cite{Haxton08-ApJ,Serenelli11-ApJ}. However, such an approach presupposes that the rate of the Bethe-Weizsäcker cycle is known with $\sim$5\% precision, better than the present 8\% \cite{Adelberger11-RMP}.

The latest comprehensive $^{14}\textrm{N(p,}\gamma\textrm{)}^{15}\textrm{O}$ experiment covering a wide energy range has been reported in 1987 by the Bochum group \cite{Schroeder87-NPA}. However, it is by now accepted that the Bochum-based value \cite{Schroeder87-NPA,Adelberger98-RMP,NACRE99-NPA} of the stellar $^{14}\textrm{N(p,}\gamma\textrm{)}^{15}\textrm{O}$ rate must be revised downward by a factor of two \cite{Adelberger11-RMP}. 

This consensus \cite{Adelberger11-RMP} is based on indirect data \cite{Bertone01-PRL,Mukhamedzhanov03-PRC,Yamada04-PLB}, direct cross section measurements \cite{Formicola04-PLB,Imbriani05-EPJA,Runkle05-PRL,Marta08-PRC,Marta11-PRC,Li16-PRC}, and R-matrix fits \cite{Angulo01-NPA,Mukhamedzhanov03-PRC}. The most important conclusion from these works is that the astrophysical S-factor, extrapolated to zero energy, for the transition to the ground state in $^{15}$O is $S_{\rm GS}(0)$ = 0.20-0.49\,keV\,barn \cite{Bertone01-PRL,Angulo01-NPA,Mukhamedzhanov03-PRC,Yamada04-PLB,Formicola04-PLB,Imbriani05-EPJA,Runkle05-PRL,Marta08-PRC,Marta11-PRC,Li16-PRC,Michelagnoli13-PhD}, not 1.55\,keV barn as previously reported \cite{Schroeder87-NPA}. 

\begin{table*}[tb]
\begin{tabular}{|l|c|c|c|c|c|}
Transition & Bochum \cite{Schroeder87-NPA} & LUNA \cite{Formicola04-PLB,Imbriani05-EPJA,Marta08-PRC,Marta11-PRC} & TUNL \cite{Runkle05-PRL} & SFII \cite{Adelberger11-RMP} & Notre Dame \cite{Li16-PRC} \\ \hline
R/DC$\rightarrow$6.79 & 1.41$\pm$0.02 & 	1.20$\pm$0.05	&	1.15$\pm$0.05	&	1.18$\pm$0.05	&	1.29$\pm$0.04(stat)$\pm$0.09(syst)	\\
R/DC$\rightarrow$6.18 & 0.14$\pm$0.05 & 	0.08$\pm$0.03	&	0.04$\pm$0.01	&	0.13$\pm$0.06	&	\\
R/DC$\rightarrow$5.24 & 0.018$\pm$0.003 & 	0.070$\pm$0.003	&					&	0.070$\pm$0.003	&	\\
R/DC$\rightarrow$5.18 & 0.014$\pm$0.004 & 	0.010$\pm$0.003	&					&	0.010$\pm$0.003	&	\\
R/DC$\rightarrow$0 & 1.55$\pm$0.34 & 		0.20$\pm$0.05	&	0.49$\pm$0.08	&	0.27$\pm$0.05	&	0.42$\pm$0.04(stat)$^{+0.09}_{-0.19}$(syst)\\ \hline 
Sum 		& 3.20$\pm$0.54	& 				1.56$\pm$0.08	&	1.68$\pm$0.09	&	1.66$\pm$0.12	&	\\ \hline \hline
\end{tabular}
\caption{\label{tab:S0} Astrophysical S-factor, extrapolated to zero energy, for the most important transitions in the $^{14}\textrm{N(p,}\gamma\textrm{)}^{15}\textrm{O}$ reaction.}
\end{table*}
\begin{figure}[b]
\includegraphics[width=\columnwidth]{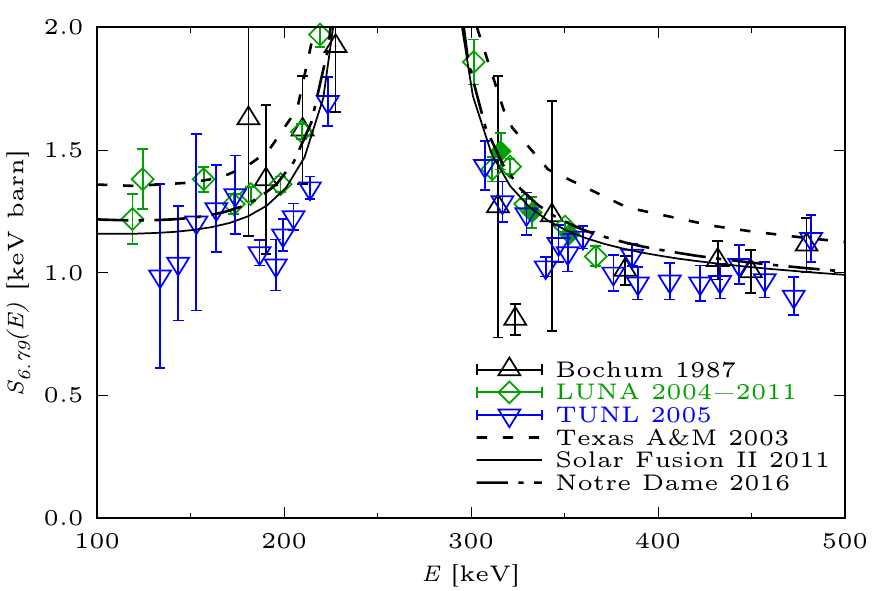}
	\caption{\label{fig:n14pg_679lo} Astrophysical S-factor for capture to the 6.79~MeV excited state in the $^{14}\textrm{N(p,}\gamma\textrm{)}^{15}\textrm{O}$ reaction at low energy from the Bochum \cite{Schroeder87-NPA}, LUNA \cite{Formicola04-PLB,Imbriani05-EPJA,Marta08-PRC,Marta11-PRC}, and TUNL \cite{Runkle05-PRL} experiments, respectively. R-matrix fits by the Texas A\&M \cite{Mukhamedzhanov03-PRC}, SFII \cite{Adelberger11-RMP}, and Notre Dame \cite{Li16-PRC} groups, respectively, are shown as lines.}
\end{figure}

The present work, instead, concentrates on the $^{14}\textrm{N(p,}\gamma\textrm{)}^{15}\textrm{O}$ transition that dominates: capture to the 6.79~MeV excited state. In addition, it also provides some new results for ground state capture. The 6.79~MeV transition accounts for $\sim$70\% of the total cross section. Its S-factor curve is essentially flat over a wide energy range \cite{Schroeder87-NPA}, indicating a dominance of direct capture and capture through very wide resonances. Indeed, the 6.79~MeV transition plays only a secondary role for the low-energy resonance at $E$ = 259 keV \cite{Marta08-PRC,Marta11-PRC}, which has recently emerged as a precise normalization point \cite{Imbriani05-EPJA,Runkle05-PRL,Bemmerer06-NPA,Adelberger11-RMP,Daigle16-PRC}. The transition has not even been detected in the subsequent resonance at $E$ = 987\,keV \cite{Marta10-PRC}. 

Several recent R-matrix extrapolations for capture to the 6.79~MeV state converge in a narrow band at  $S_{\rm 6.79}(0)$=1.15-1.20~keV~barn, with error bars as low as 4\% \cite{Formicola04-PLB,Imbriani05-EPJA,Runkle05-PRL,Adelberger11-RMP}. Two works, however, report somewhat higher central values and also higher uncertainties. The first, based on a measurement of the asymptotic normalization coefficient governing direct capture and a subsequent R-matrix fit including the data available at the time (i.e. without the LUNA and TUNL data), reported $S_{\rm 6.79}(0)$~=~(1.40$\pm$0.20)~keV~barn \cite{Mukhamedzhanov03-PRC}. The second, based on a comprehensive R-matrix fit including not only new capture data but also angular distributions, gave a value of $S_{\rm 6.79}(0)$~=~(1.29$\pm$0.04(stat)$\pm$0.09(syst))~keV~barn \cite{Li16-PRC}. 

These various R-matrix fits may be benchmarked against recent and precise experimental capture data at relatively low energy, $E$ = 100-500~keV (Fig.~\ref{fig:n14pg_679lo}). However, it should be noted that there is still a significant energy gap from the data points at 100-500~keV to the solar Gamow energy, $E_{\rm Gamow}$~=~27~keV. 
Summing detector data from LUNA reach down to the lowest energies hitherto measured, $E$~=~70~keV, and provide a value for the total S-factor, summed from all transitions, of $S(E=70\,{\rm keV})$=1.74$\pm$0.14$_{\rm stat}$$\pm$0.14$_{\rm syst}$~keV~barn \cite{Lemut06-PLB,Bemmerer06-NPA}. However, by design the summing data cannot constrain the partial S-factor for capture to the 6.79~MeV level very well. 

Even though the experimental situation at $E$ = 100-500~keV is satisfactory (Fig.~\ref{fig:n14pg_679lo}), for several important energy intervals at higher energy, $E$~$>$~500~keV, the only existent radiative capture data set is still the one from Bochum \cite{Schroeder87-NPA}. As mentioned earlier, for another transition in $^{14}\textrm{N(p,}\gamma\textrm{)}^{15}\textrm{O}$ , ground state capture, the Bochum data had to be corrected by up to 50\% \cite{Formicola04-PLB,Adelberger11-RMP} for the so-called true coincidence summing-in effect \cite{Gilmore08-Book}. This effect led to an artificial increase of the signal for ground state capture by the coincident detection of the DC$\rightarrow$6.79 and 6.79$\rightarrow$0 $\gamma$-rays. It was neglected in the original publication \cite{Schroeder87-NPA} but corrected for in subsequent work \cite{Formicola04-PLB,Adelberger11-RMP}. The same process leads to the loss of counts in the 6.79$\rightarrow$0 $\gamma$-ray, by the true coincidence summing-out effect. This latter effect scales with the total $\gamma$-ray detection efficiency and may thus reach values up to 10\% in close geometry. The Bochum excitation function was taken at close distance, with just 2\,cm separating the target from the detector endcap \cite{Schroeder87-NPA}.

\begin{figure*}[tb]
\includegraphics[width=1.5\columnwidth]{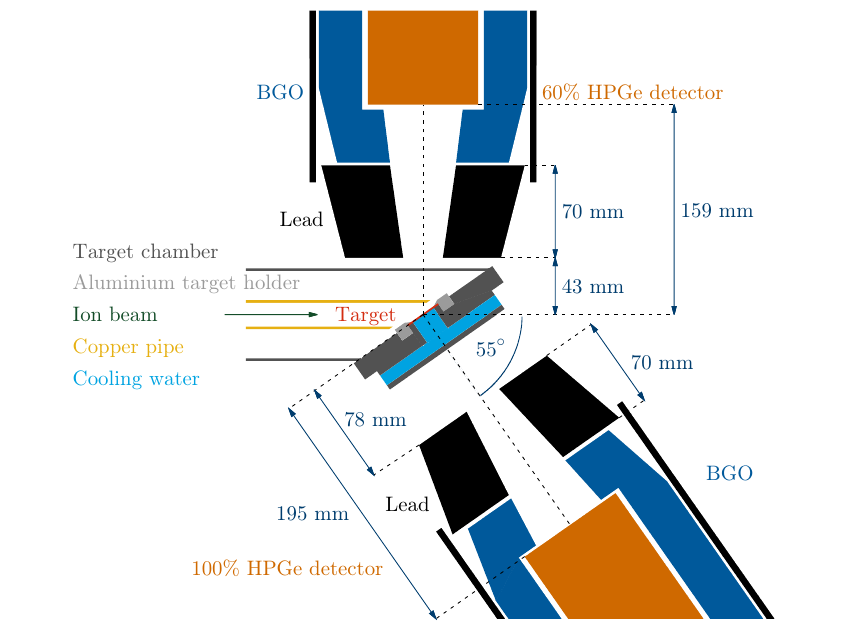}
\caption{\label{fig:targetchamber}Schematic top view of the experimental setup. The $^1$H$^+$ beam impinges from the left side. The HPGe crystals in their end caps shown in orange are surrounded by BGO scintillators in blue and lead shielding in black.}
\end{figure*}

A very recent study from Notre Dame \cite{Li16-PRC} contributed radiative capture data for capture to the 6.79~MeV level in the 1.5-3.4~MeV energy range and for the ground state transition from 0.6-3.4~MeV. In the important energy range from 0.5-1.5~MeV, for the 6.79~MeV transition, angular distributions are reported but integrated S-factor data are missing \cite{Li16-PRC}.

Because of the possible systematic uncertainty given by unaccounted for summing effects in the Bochum data set \cite{Schroeder87-NPA} and the limitations of the very recent Notre Dame data set, it is necessary to perform an independent experimental study of the $^{14}\textrm{N(p,}\gamma\textrm{)}^{15}\textrm{O}$ cross section over a wide energy range. The present work aims to provide this independent cross-check, by supplying new and independent capture cross section data for two $^{14}\textrm{N(p,}\gamma\textrm{)}^{15}\textrm{O}$ transitions in the $E$ = 366-1289\,keV energy range. The energy range is chosen such that there is some overlap to the recent and precise low-energy data at 100-500\,keV from LUNA and TUNL \cite{Formicola04-PLB,Imbriani05-EPJA,Runkle05-PRL,Marta08-PRC,Marta11-PRC}.

This work is organized as follows. The experimental setup is described in \Sec{sec:setup}. \Sec{sec:measurements} discusses the irradiations and analyses performed. The experimental results are shown and discussed in \Sec{sec:results}, and \Sec{sec:rmatrix} describes an R-matrix fit including the new data. A summary and an outlook are offered in \Sec{sec:summary}.


\section{\label{sec:setup}Experimental setup}

The 3\,MV high-current Cockroft-Walton tandem accelerator \cite{Friedrich96-NIMA} at Helmholtz-Zentrum Dresden-Rossendorf (HZDR) provided a proton beam with energies of $E_{\rm p}$ = 400-1400~keV \footnote{In this work, $E_{\rm p}$ is used to denote the projectile energy in the laboratory system}. The H$^-$ beam generated by an IONEX 860C cesium sputter ion source from TiH$_2$ sputter targets was magnetically analyzed, then sent into the tandem accelerator, where it was accelerated, stripped to H$^+$ on the high voltage terminal, and further accelerated. 

The beam then passed a switching magnet, electrostatic deflector panels and a neutral particle trap before reaching the final beam limiting collimator. At this collimator with 5\,mm diameter, at least 10\% of the beam intensity was deposited in order to ensure a homogeneous beam on target.

The acceleration voltage was measured with a precision voltage divider that was read out with a 3.5 digit digital multimeter. The readout chain was calibrated using sharp (p,$\gamma$) resonances in the energy range studied here, giving a final uncertainty of the ion beam energy of better than 1\,keV, with 0.5\,keV reproducibility. 

\subsection{\label{subsec:chamber}Target chamber}

The target chamber (\fig{fig:targetchamber}) has been adapted from previous experiments \cite{Marta10-PRC,Schmidt13-PRC,Schmidt14-PRC,Depalo15-PRC,Reinhardt16-NIMB} with minor modifications. At a distance of 50\,cm downstream of the final collimator, solid targets were placed with their normal at an angle of 55$^\circ$ with respect to the beam direction. 

The target chamber and beam lines were evacuated by turbomolecular pumps backed by a scroll pump and rotary vane pumps for the target chamber and the beam lines, respectively. Rubber-free Viton seals were used throughout. The typical pressure in the target chamber and also in the beam line was 3$\times 10^{-7}$~mbar with the beam on target. 

The $^1$H$^+$ beam current was 3-16\,$\mu$A. The lowest beam energies used here required a terminal voltage below 0.2\,MV, resulting in limited transmission of the 3\,MV tandem and limiting the beam current to 3-5\,\,$\mu$A. At the highest energies used here, 0.7\,MV terminal voltage was used and transmission was excellent, leading to 16\,$\mu$A beam current on target.
 
In order to dissipate the heat during the irradiation of the targets, the 0.22\,mm thick tantalum target backing was directly water cooled. A 13\,cm long, -100~V biased copper pipe of 2.2\,cm diameter extended to 0.2\,cm distance from the target surface and suppressed secondary electrons emitted from the target. The electrical current from the target was measured with an Ortec model 439 Digital Current Integrator and recorded both with a scaler and in the list mode data acquisition. The precision of the beam current calibration was estimated as 1\%.

\subsection{\label{subsec:target} Targets}

Titanium nitride was selected as target material, because this ceramic material has shown both a favorable stoichiometric ratio near Ti$_1$N$_1$ and excellent stability under ion bombardment in previous studies \cite{Imbriani05-EPJA,Marta08-PRC,Marta10-PRC,Marta11-PRC,Reinhardt16-NIMB}. Standard 0.22\,mm thick tantalum disks of 27\,mm diameter were used as target backing, allowing the backing to be in direct contact with cooling water.

The TiN targets were produced by the reactive sputtering technique \cite{Rigato01-SCT}. Three of the five targets used here were produced at INFN Padova, Italy and the other two at HZDR. The TiN layer was 140-400\,nm thick, with a stoichiometric ratio Ti:N approaching 1:1. Relevant details of the targets are summarized in Table~\ref{tab:targets}.

\begin{table}[tb]
\begin{tabular}{|l|r|l|p{35mm}|} \hline
Designator & $d$ [nm] & Stoich. & $E_{\rm p}$ [keV] \\ \hline \hline
Vo-TiN-5 & 170 & TiN$_{0.81}$  & 747, 856, 957\\
Vo-TiN-6 & 170 & TiN$_{0.80}$  & 533, 1115, 1191, 1301\\
St-TiN-1 & 140 & TiN$_{0.83}$  & 640 \\ 
St-TiN-5 & 170 & TiN$_{0.87}$  & 407, 852 \\
Ca-TiN-2 & 200 & TiN$_{0.97}$  & 640, 681, 852, 1401\\ \hline 
\end{tabular}
\caption{\label{tab:targets} Titanium nitride targets used, their nominal thickness $d$, and Ti:N stoichiometric ratio as determined by the elastic recoil detection method (\Sec{subsec:erda}). The beam energies for which the targets were used are given, as well. }
\end{table}

\subsection{\label{subsec:gammadetectors} $\gamma$-ray detectors}
\label{subsec:gammadetectors}

The $\gamma$-ray detection setup consisted of two high purity germanium (HPGe) detectors with 60\% and 100\% relative efficiency, respectively. Each of the two HPGe detectors was surrounded by a BGO scintillator (minimum thickness 3\,cm) for escape suppression and lead (1\,cm thickness) for shielding against background radiation. 

The lead collimator of detector 1 (100\% efficiency) was placed at 55$^\circ$ with respect to the beam axis, directly behind the TiN target (\fig{fig:targetchamber}), at 195\,mm distance to the target. Detector 2 was placed at 90$^\circ$ angle and 159\,mm distance to the target. This second detector helped to place a limit on possible angular distribution effects and increased the solid angle covered and thus the statistics. 

\begin{figure}[t]
\includegraphics[width=0.99\columnwidth]{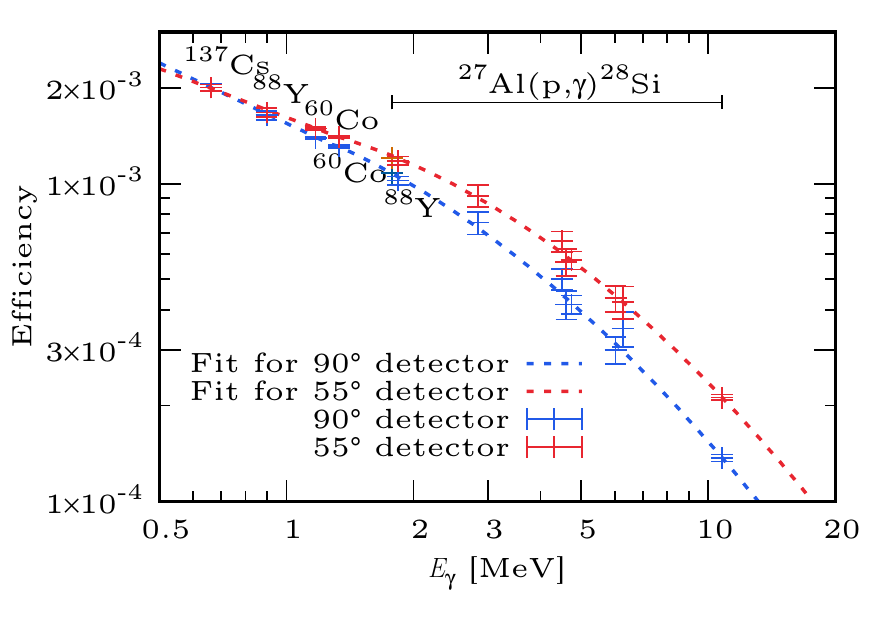}
\caption{\label{fig:effic} The calculated $\gamma$ detection efficiency for detector 1 at 55$^\circ$ (red dots) and detector 2 at 90$^\circ$ (blue dots) tagged with the source (decay or $^{27}$Al+p reaction) and the fitted efficiency curves for both detectors (red and blue lines)}
\end{figure}

For the determination of the full-energy peak detection efficiency as a function of energy, $^{60}$Co, $^{88}$Y, and $^{137}$Cs $\gamma$-ray intensity standards calibrated to activity uncertainties better than 1\% (68\% confidence level) by Physikalisch-Technische Bundesanstalt (PTB), Braunschweig, Germany were used. In addition, the well known \Alpg reaction \cite{Anttila77-NIM} was used to extend the efficiency curve up to 11\,MeV (\fig{fig:effic}). Henceforth, an empirical parameterization of the detection efficiency curve (lines in \fig{fig:effic}) was used, assuming 3\% uncertainty.

In addition, the target chamber and $\gamma$-ray detectors were modelled in the GEANT4 \cite{Agostinelli03-NIMA} Monte Carlo framework using the nominal geometry provided by the manufacturer. The Monte Carlo simulation was only used for the prediction of the shape of the Compton edge and continuum. The peak detection efficiency for the data analysis was always taken from the experimental data and their parameterization (\fig{fig:effic}) instead.

\section{\label{sec:measurements}Measurements}

Due to the low absolute value of the $^{14}\textrm{N(p,}\gamma\textrm{)}^{15}\textrm{O}$ cross section, the measurements entailed long irradiations lasting between 3 and 97 hours. Thus it was essential to ensure the stability of the relevant experimental conditions. 

The irradiations at the energy to be studied were bracketed by target studies by nuclear resonant reaction analysis (NRRA), performed in situ by tuning the beam energy to that of the 897\,keV resonance in the $^{15}\textrm{N(p,}\alpha\gamma\textrm{)}^{12}\textrm{C}$ reaction (\Sec{subsec:nrra}). In case a data point required more than 24 hours of irradiation, an NRRA run was interjected daily. In addition, each target was analyzed by the Elastic Recoil Detection (ERD) technique, after the irradiations and in a different setup (\Sec{subsec:erda}). During the irradiations themselves, the yield of the strong 4.4\,MeV $\gamma$-ray from the non-resonant $^{15}$N(p,$\alpha\gamma$)$^{12}$C reaction was continually used to monitor the stability of the targets (\Sec{subsec:irradiations}).

\subsection{\label{subsec:nrra}Nuclear resonant reaction analysis (NRRA)}

\begin{figure}[t]
\includegraphics[width=0.99\columnwidth,trim=4mm 0 0 0,clip]{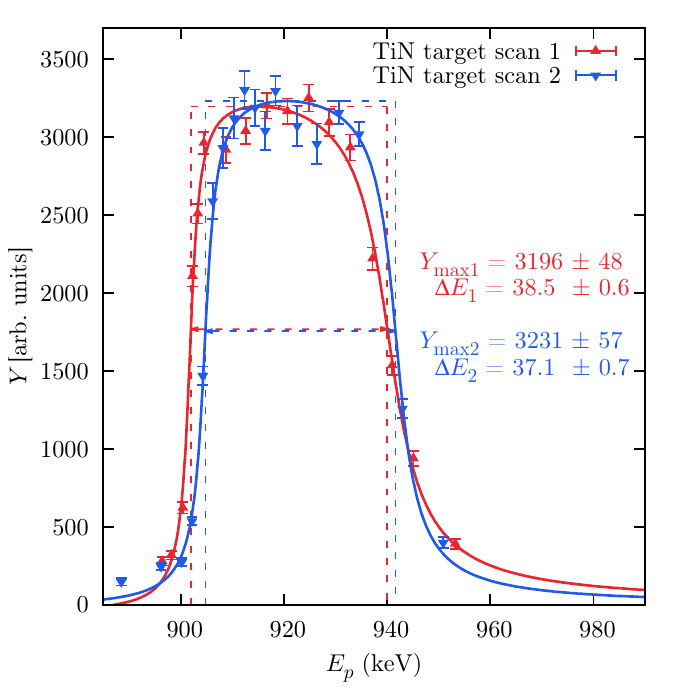}
\caption{\label{fig:tscan} Resonance scans (in arbitrary units) of target Ca-TiN-2 before the first (red) and after the last (blue) irradiation of this sample, total integrated charge on target 1.92 C. The 897\,keV resonance in the $^{15}\textrm{N(p,}\alpha\gamma\textrm{)}^{12}\textrm{C}$ reaction was scanned by tuning the proton beam energy $E_\textrm{p}$ and plot it against the yield (in arbitrary units) of the 4.4\,MeV $\gamma$-ray on the y-axis. See text for details.}
\end{figure}

For the in situ analysis of the targets, the $E_{\rm p}$ = 897\,keV resonance in the $^{15}$N(p,$\alpha\gamma$)$^{12}$C reaction was selected. This resonance is very strong, $\omega\gamma$ = 362$\pm$20\,eV \cite{Marta10-PRC}, indeed so strong that the low isotopic abundance (0.3663\% \cite{Coplen02-PAC}) of $^{15}$N in the natural nitrogen used for TiN production is compensated and that this is the strongest resonance available in the proton beam energy range used here. Thus no great changes in focusing were needed when switching from the irradiation to the NRRA run and vice versa.

The strongly anisotropic $\gamma$-ray angular distribution of this resonance and a lower-energy, weaker one at $E_{\rm p}$ = 430\,keV has last been studied in details in the 1950s \cite{Barnes52-CJP,Kraus53-PR}. A very recent re-study of the $E_{\rm p}$ = 430\,keV resonance's $\gamma$-ray angular distribution led to somewhat different results \cite{Reinhardt16-NIMB}. Therefore, pending further re-investigation also of the 897\,keV resonance, in the present work the NRRA results obtained with this latter resonance are used only for relative monitoring of one given target between the start and the end of the irradiation. 

During an NRRA run, the beam energy was tuned in 1-5\,keV steps, and the yield of the broad 4.4\,MeV $\gamma$-ray from the decay of the first excited state of $^{12}$C was plotted, leading to a precise profile allowing to judge both the width and the distribution of the nitrogen in the target (\fig{fig:tscan}). 

The NRRA scans showed the targets to be stable under bombardment. However, in several cases a buildup of a layer on top of the target was observed, resulting in a slight shift of the resonance profile to higher beam energies (\fig{fig:tscan}). This layer was also apparent after the irradiations as a slight darkening of the beam spot. For practical reasons, no liquid nitrogen cooled cold trap was used here. Apparently the observed target chamber pressure of 3$\times$10$^{-7}$\,mbar was not low enough to entirely prevent the buildup of a parasitic layer. The proton beam energy loss in this layer was 
taken into account in the analysis (see below, Section~\ref{subsec:sfactor}), based on the resonance scans.

In addition to the plateau yield, which is used to monitor the relative stoichiometry of the target, also the energetic target thickness $\Delta E_p^{897}$ has been determined from the NRRA scans of each target used here, both before and after the irradiation at the energy under study.

\subsection{\label{subsec:erda}Elastic Recoil Detection Analysis (ERDA)}

\begin{figure}[t]
\includegraphics[angle=-90,width=0.99\columnwidth]{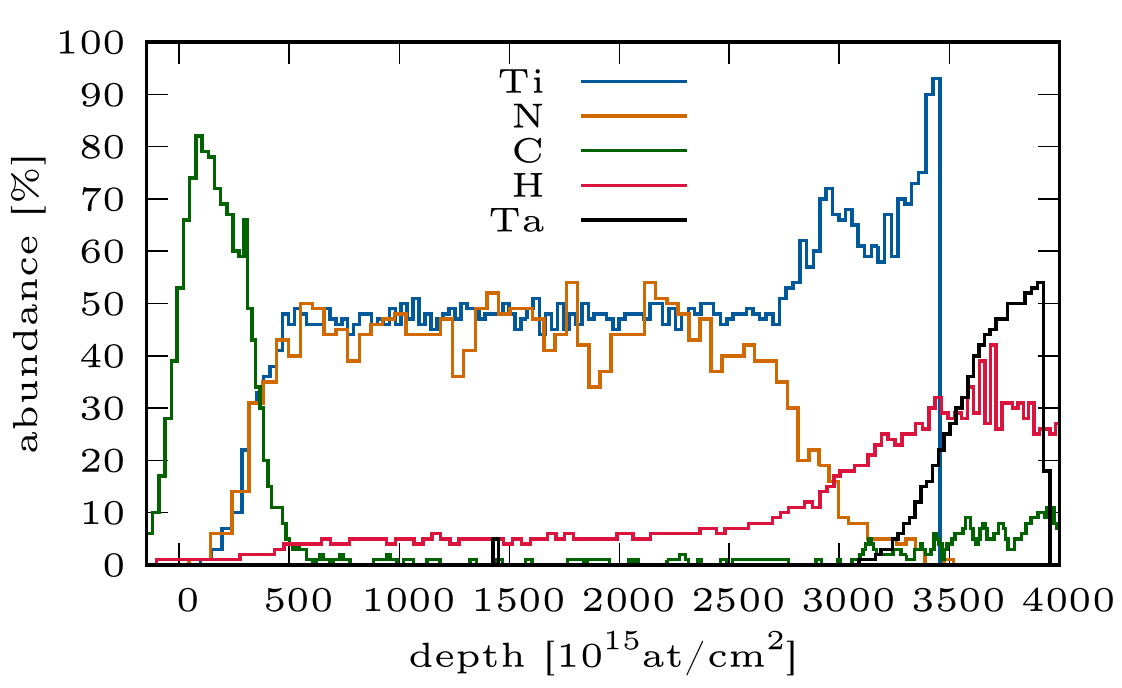}
\caption{\label{fig:profile} Depth profile of target Ca-TiN-2 in the beam spot area. See text for details.}
\end{figure}

After the irradiations concluded, each target was analyzed by the heavy-ion elastic recoil detection (HI-ERD) technique at the HZDR 6\,MV Tandetron accelerator. For the analysis, 43 MeV chlorine ions were used. The HZDR HI-ERD setup and analysis techniques have been described previously \cite{Kosmata14-NIMB}. For further analysis, the ERD data were converted to a depth profile using the NDF software \cite{Barradas97-APL}.

For each target used here, a point well inside the beam spot area was studied by the HI-ERD technique. In addition, for a number of targets also a second spot well outside the beam spot was studied by HI-ERD, in order to independently verify the degradation of the target under bombardment, in addition to the resonance scans (\Sec{subsec:nrra}).

The data are discussed using the results from target Ca-TiN-2 (\fig{fig:profile}). The initial layer found already in the NRAA scan is reproduced by ERD, and it is found to be carbon. This is consistent with the fact that the beam spot appears slightly blackened. The integral thickness of the carbon layer from HI-ERD is 270$\times$10$^{15}$ atoms/cm$^2$, which leads to a proton energy loss of 2.7\,keV at the lowest beam energy this particular target was used for, $E_p$ = 640\,keV. This number is consistent with the shift determined from the two 897\,keV NRRA scans of this target (Section~\ref{subsec:nrra}), which result in 2.2\,keV energy loss at $E_p$ = 640\,keV.  

Below the thin carbon layer, a 2590$\times$10$^{15}$ atoms/cm$^2$ thick layer of TiN is found, with a stoichiometric ratio of TiN$_{0.97}$. This layer also contains 5-10 atom\% hydrogen. This element is usually found in tantalum and may have migrated from the backing, where the ERD analysis shows a peak in the hydrogen concentration up to 30 atom\%, to the TiN layer. Behind the TiN layer, a thin Ti layer is present that may have been created due to a delayed ignition of the plasma, leading to some pure Ti to be evaporated on the Ta backing. In this layer a small oxygen contamination ($<$8\%, not shown in the plot to avoid confusion) was found that indicates some oxidation of the backing. Finally, as expected, towards the end of the TiN target a steep increase of the concentration of the backing material, tantalum, is found.

Separately for each target, the stoichiometric ratio $x$, for a compound TiN$_x$, was determined by fitting the titanium and nitrogen concentrations found by HI-ERD on their common plateau (Table~\ref{tab:targets}). Using the value $x$ thus obtained, the effective stopping power \cite{Iliadis15-Book} $\epsilon_{\rm eff}^{14}$(897) for protons with $^{14}$N as the active nucleus was determined:
\begin{equation}\label{eq:effstop}
\epsilon_{\rm eff}^{14}(897) = \frac{1}{0.996337} \left[ \epsilon_{\rm N}(897) + \frac{1}{x}\epsilon_{\rm Ti}(897) \right]
\end{equation}
where $\epsilon_{\rm N}$(897) and $\epsilon_{\rm Ti}$(897) are the stopping powers for 897\,keV protons in solid nitrogen and titanium, respectively. The stopping power values $\epsilon_{\rm N, Ti}$ are taken from the SRIM software \cite{Ziegler10-NIMB}, adopting the SRIM relative uncertainty of 2.9\% for $\epsilon_{\rm N}$ and 4.4\% for $\epsilon_{\rm Ti}$. The factor 1/0.996337 corrects the effective stopping power for the 99.6337\% isotopic abundance of $^{14}$N in natural nitrogen. This abundance has been found to be very stable in air \cite{Coplen02-PAC} and is conservatively assumed to hold to within 1\% here.


\subsection{\label{subsec:irradiations}Irradiations}

\begin{figure}[t]
\includegraphics[angle=-90,width=\columnwidth]{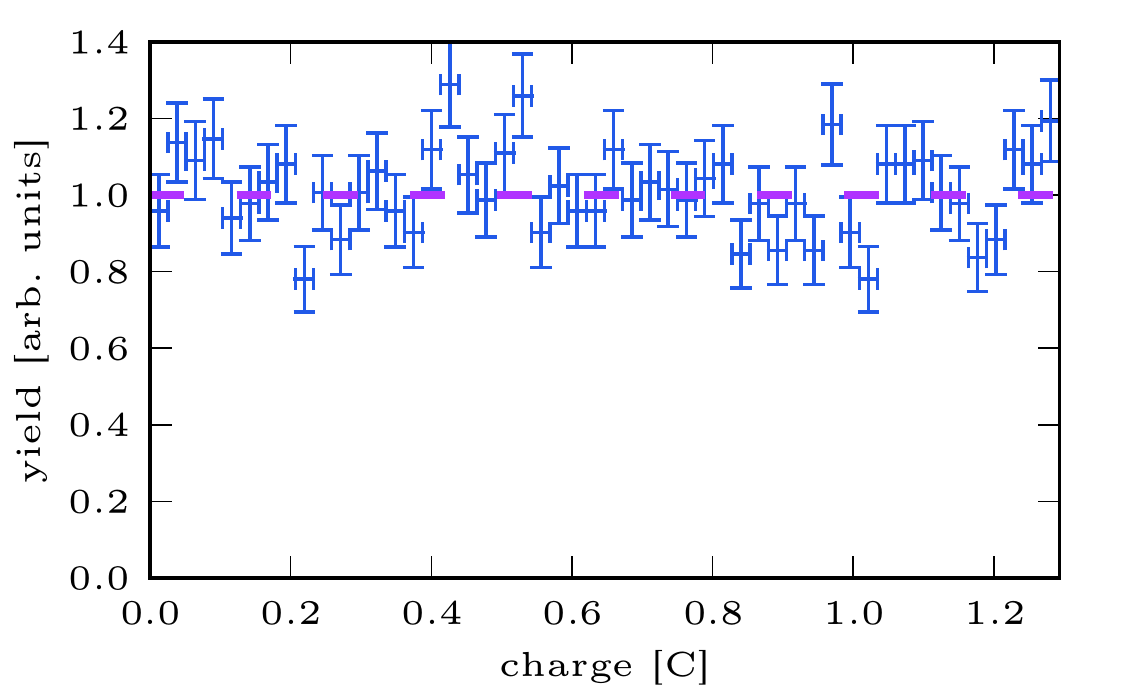}
\caption{\label{fig:yield} Yield of the 4439\,keV $\gamma$-ray from the $^{15}$N(p,$\alpha\gamma$)$^{12}$C reaction during a long irradiation of target St-TiN-5 at $E_{\rm p}$ = 407 keV, as a function of the accumulated charge. }
\end{figure}

The $^{14}\textrm{N(p,}\gamma\textrm{)}^{15}\textrm{O}$ reaction was studied at twelve proton beam energies between 0.4-1.4\,MeV, selected to avoid as much as possible parasitic resonances as well as the $E_{\rm p}$ = 1058\,keV resonance in $^{14}\textrm{N(p,}\gamma\textrm{)}^{15}\textrm{O}$.

During an irradiation, instead of the very low yield of the reaction under study, the much more probable non-resonant $^{15}$N(p,$\alpha\gamma$)$^{12}$C reaction was used as a monitor. The yield of the 4439\,keV $\gamma$-ray from the decay of the first excited state of $^{12}$C provided a real time estimate of the state of each target.

The only case where a significant degradation of the 4439\,keV yield was observed was target St-TiN-1, with 28\% degradation. This target was then excluded from the analysis to limit the resultant uncertainty. Only runs with 4439\,keV yield degradation less than 5\% were adopted for the analysis.

In addition to this yield monitoring, also the target and collimator currents were regularly recorded. In case of a significant deterioration of the target current or of the target/collimator current ratio, the ion source and beam transmission were re-adjusted. 

\begin{figure*}[t]
\includegraphics[width=\textwidth]{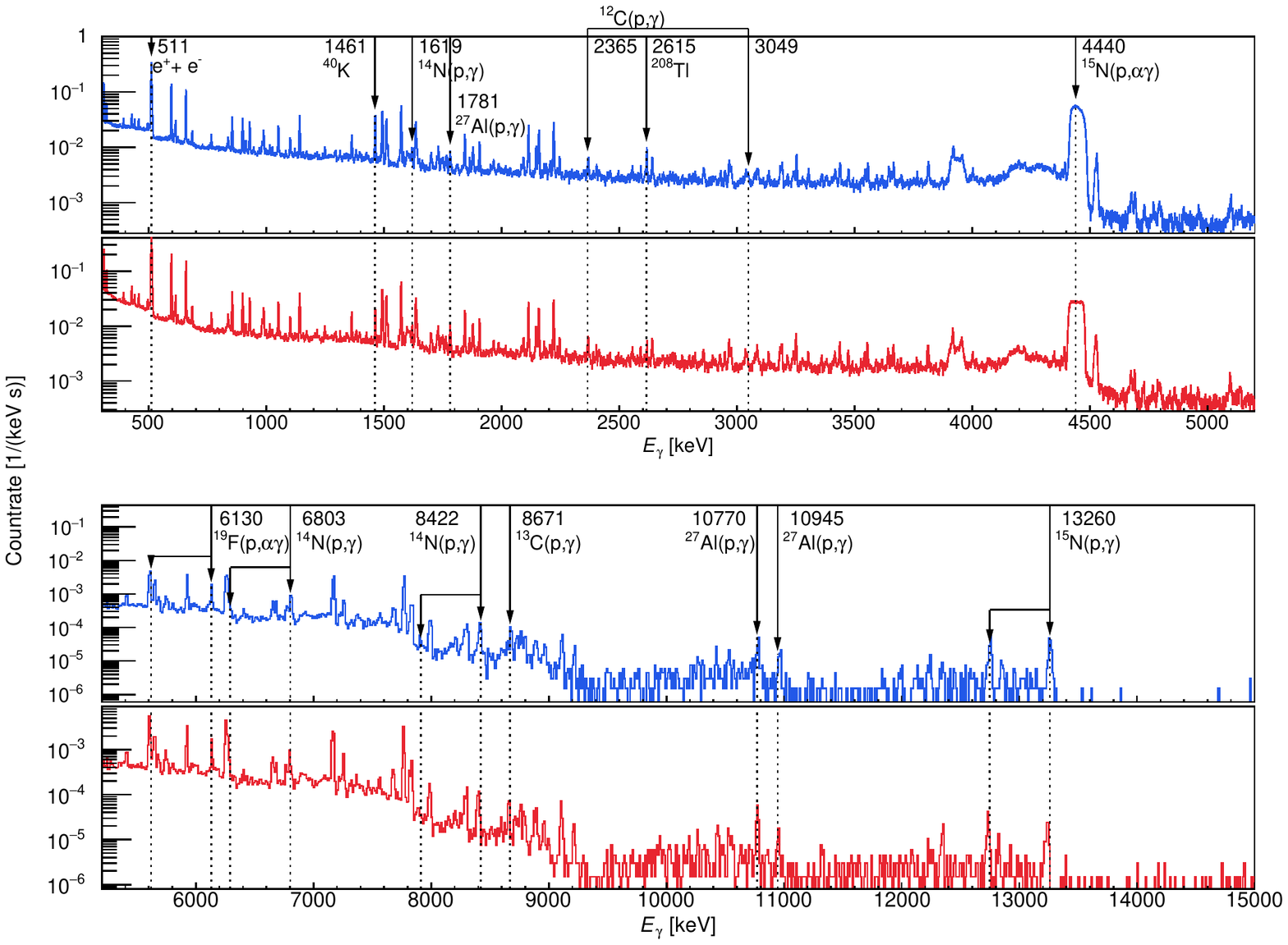}
\caption{\label{fig:spec1191} 
In-beam $\gamma$-ray spectrum at $E_p$ = 1191\,keV. 
The irradiation time was 3.7 hours, total accumulated charge 0.21 C. 
Top, blue spectrum: 55$^\circ$ detector. Bottom, red spectrum: 90$^\circ$ detector.
}
\end{figure*}

\begin{figure*}[t]
\includegraphics[width=\textwidth]{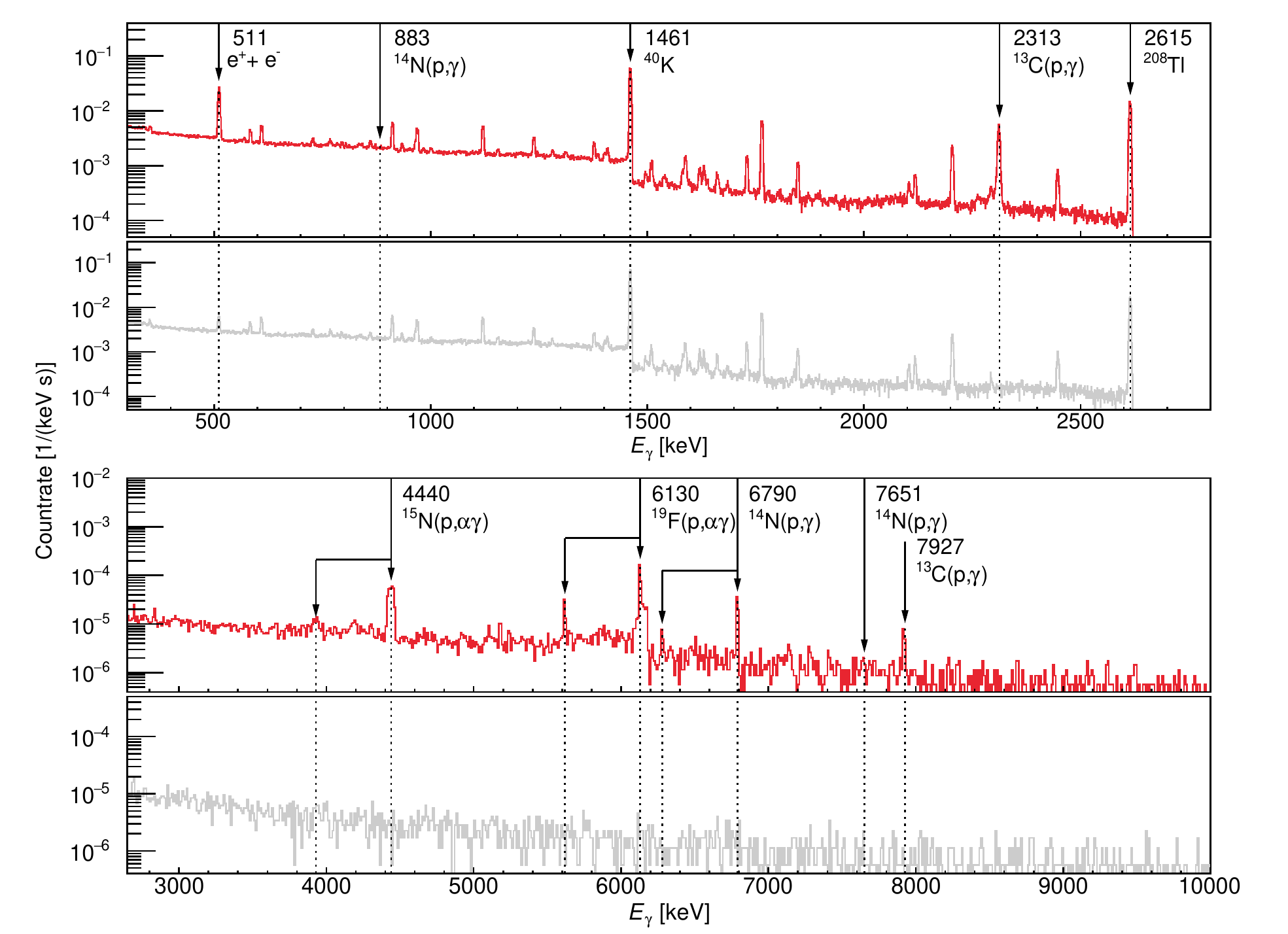}
\caption{\label{fig:spec407} 
In-beam $\gamma$-ray spectrum at $E_p$ = 407\,keV, in the 90$^\circ$ detector. The top, red spectrum is the in-beam spectrum with an irradiation time of 97 hours and a total accumulated charge of 1.3 C. The bottom, grey spectrum is the no-beam background, rescaled for equal time.
}
\end{figure*}

\section{\label{sec:results}Data Analysis and Results}

For the data analysis, in a first step the recorded in-beam $\gamma$-ray spectra are interpreted (\Sec{subsec:gammaspectra}). Then, the efficiency-corrected yields from the two detectors are compared to check the literature data on the angular distribution (\Sec{subsec:angdist}). The sought after cross section and astrophysical S-factor are determined from the yields and angular corrections (\Sec{subsec:sfactor}). The uncertainties are summarized (\Sec{subsec:uncertainties}), and the results are discussed (\Sec{subsec:Discussion}).

\subsection{\label{subsec:gammaspectra}Interpretation of the observed $\gamma$-ray spectra}

Typical in-beam $\gamma$-ray spectra taken with the two HPGe detectors are shown in \fig{fig:spec1191} for a representative high beam energy, $E_p$ = 1191\,keV, and in \fig{fig:spec407} for the lowest beam energy, $E_p$ = 407\,keV. 

In the low $\gamma$-ray energy part of the 1191\,keV spectrum (\fig{fig:spec1191}, upper panel), the well-known room background lines at 511, 1461, and 2615\,keV are visible. In addition, the primary $\gamma$-ray from capture to the 6792\,keV level can be seen at 1619\,keV, with the typical peak shape given by the target profile. One of the tallest peaks is actually the wide 4439\,keV line from $^{15}$N(p,$\alpha\gamma$)$^{12}$C, which is used for the monitoring of the irradiation (\Sec{subsec:irradiations}). Somewhat weaker lines from the $^{12}$C(p,$\gamma$)$^{13}$N reaction on the initial carbon layer of the target are apparent at 2365 and 3049\,keV.

In the high $\gamma$-ray energy part of the spectrum (\fig{fig:spec1191}, lower panel), a number of parasitic peaks due to the $^{19}$F(p,$\alpha\gamma$)$^{16}$O background reaction are apparent, most problematic at 6130\,keV. This peak, which includes both a sharp Gaussian component due to $^{16}$O nuclei stopped in the backing and a wide Doppler continuum due to in-flight decay of $^{16}$O, is so close to the weak secondary $\gamma$-ray from the decay of the 6172\,keV excited state in $^{15}$O that no analysis of the DC$\rightarrow$6172 transition is attempted in the present work.

Additional $\gamma$-lines stem from the $^{13}$C(p,$\gamma$)$^{14}$N reaction on the initial carbon layer, from the $^{15}$N(p,$\gamma$)$^{16}$O reaction on the 0.4\% $^{15}$N content in natural nitrogen, and the $^{27}$Al(p,$\gamma$)$^{28}$Si reaction from beam losses on the target holder. This latter reaction also gives rise to a secondary $\gamma$-ray at $E_\gamma$ = 1779\,keV that is populated in a number of strong $^{27}$Al(p,$\gamma$)$^{28}$Si resonances. Based on their known strengths and branching ratios \cite{Meyer75-NPA,NDS28-2013}, this line is used to put an upper limit of $\leq$0.5\% for beam lost on the target holder.

Despite the low counting rate at $E_{p}$ $<$ 0.5\,MeV, low-energy runs were undertaken here in order to connect the present data to the well-studied low-energy region at $E$ = 300-500\,keV (\fig{fig:n14pg_679lo}). The $\gamma$-ray spectrum from the run with the lowest proton energy is shown in \fig{fig:spec407}. 

In the low $\gamma$-ray energy region, the spectrum is dominated by room background (\fig{fig:spec407}, top panel). Even the Compton continuum from the room background is so strong to prevent a meaningful analysis of the primary $\gamma$-ray from capture to the 6792\,keV level (at $E_\gamma$ = 886\,keV). The primary $\gamma$-ray from the $^{12}$C(p,$\gamma$)$^{13}$N reaction on the thin carbon layer on top of the target is apparent, as well.

The broad 4439\,keV peak by the $^{15}\mathrm{N(p,}\alpha\gamma)^{12}\mathrm{C}$ reaction is again clearly visible in the high-energy part of the spectrum (\fig{fig:spec407}, bottom panel). Of the $^{19}\mathrm{F(p,}\alpha\gamma)^{16}\mathrm{O}$ peaks, only the most problematic one at 6130\,keV is visible at this low energy.  The secondary $\gamma$-ray due to the decay of the $E_{\rm x}$ = 6792\,keV excited state in $^{15}$O (shown at 6790\,keV in \fig{fig:spec407}) is clearly visible. However, at this lowest beam energy, the same is not true for the primary $\gamma$-ray from ground state capture, expected at 7674\,keV. It coincides with the Compton edge of the 7927\,keV peak from the direct capture peak in the $^{13}\mathrm{C(p,}\gamma)^{14}\mathrm{N}$ reaction, preventing an analysis of the ground state transition for this data point.

The three $\gamma$-rays used for the analysis of the nuclear reaction of interest,  $^{14}$N(p,$\gamma$)$^{15}$O, are (1) the primary $\gamma$-ray from capture to the 6792\,keV level, (2) the secondary $\gamma$-ray due to the decay of the $E_{\rm x}$ = 6792\,keV excited state in $^{15}$O, and (3) the primary $\gamma$-ray from ground state capture. These three $\gamma$-rays are shown in details in Figures \ref{fig:multispec1} and \ref{fig:multispec2}, together with the regions of interest selected for the determination of the peak area, and for the estimation  of the linear background to be subtracted.

In several cases, special steps had to be taken for the background subtraction; they are listed in the following text. The respective error bar for each of the data points listed was increased to take the uncertainty from the subtraction procedure into account.

\begin{figure}[t]
\includegraphics[width=\columnwidth]{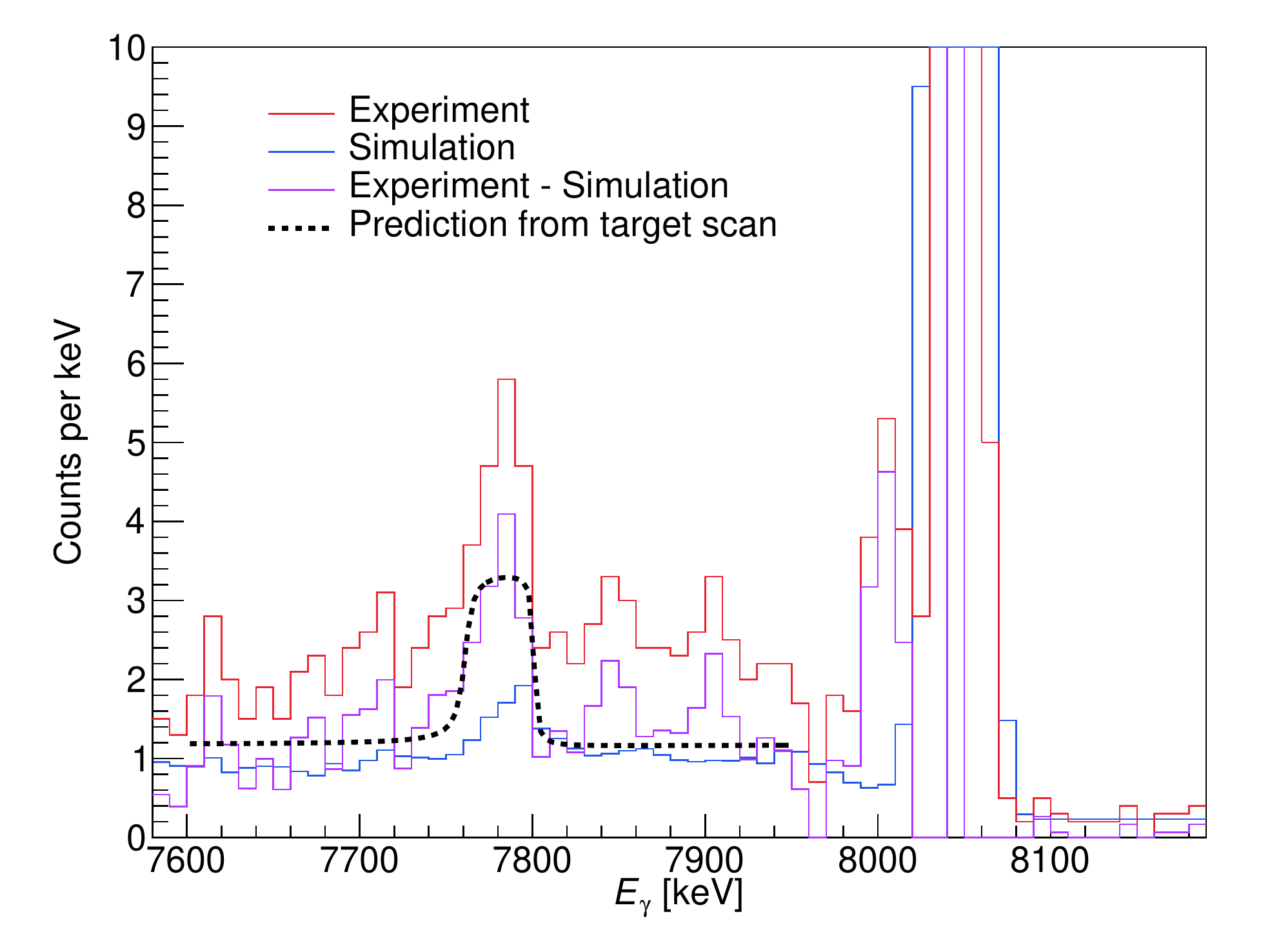}
\caption{\label{fig:spec533} 
In-beam $\gamma$-ray spectrum from the 55$^\circ$ detector for the $E_p$ = 533\,keV run near the ground state capture peak. See text for details.
}
\end{figure}

\begin{figure*}[t]
	\begin{adjustbox}{addcode={\begin{minipage}{\width}}{\caption{\label{fig:multispec1} 
		In-beam $\gamma$-ray spectra from the 55$^\circ$ detector for the $E_p$ = 407-852\,keV runs. The three columns show, from left to right, the primary $\gamma$-ray from capture to the 6792\,keV level, the 6792\,keV secondary $\gamma$-ray, and the ground state capture peak. Cases where no clean peak could be identified and that were thus not analyzed are marked with an empty panel. The regions of interest for the peak area and for the background estimation are marked. The ground state capture peak at $E_p$ = 533\,keV is a special case and discussed in the text and in \fig{fig:spec533}.
		}\end{minipage}},rotate=90,center}
		\includegraphics[angle = 0, width=\textheight]{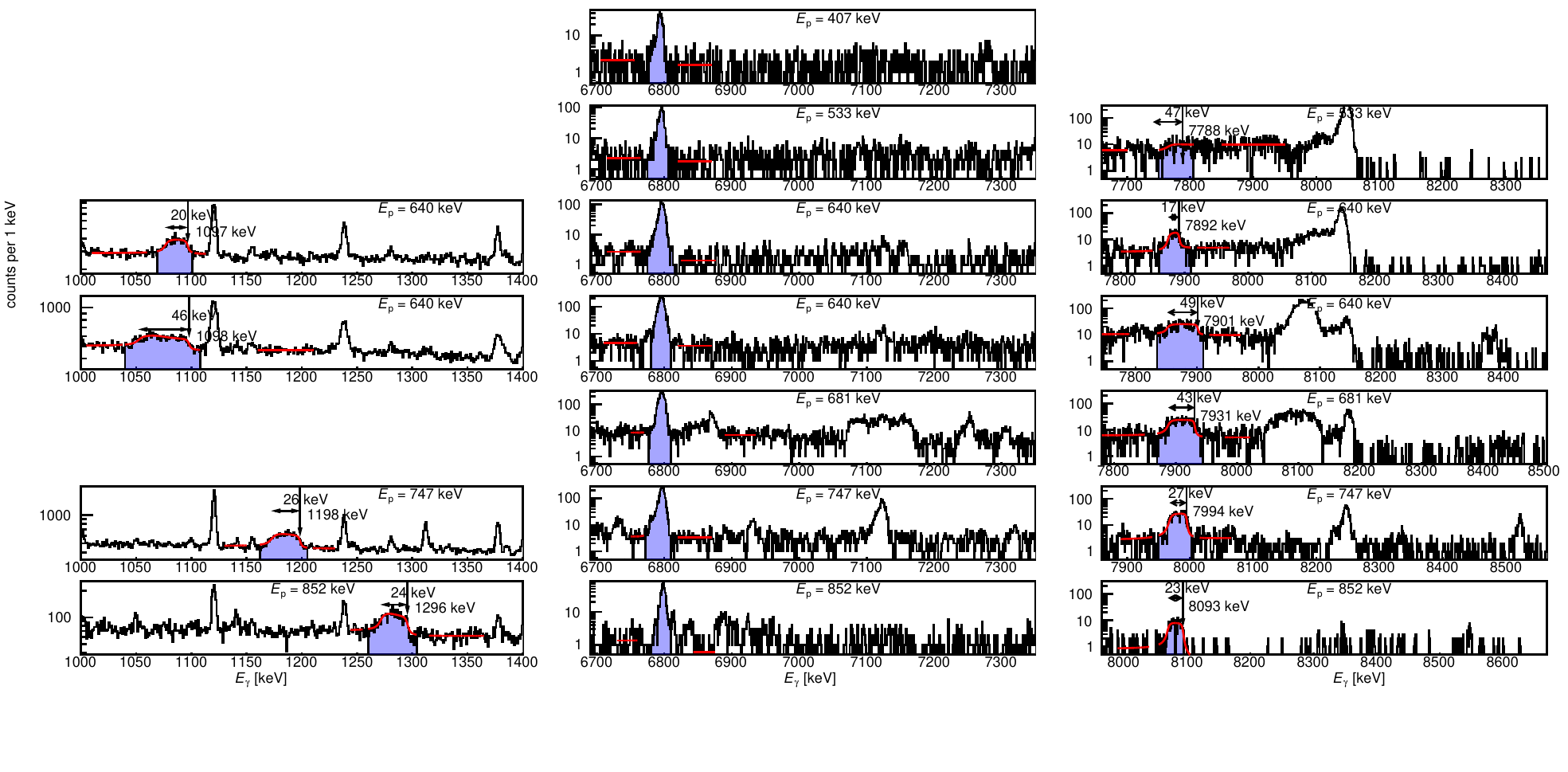}
	\end{adjustbox}
\end{figure*}

\begin{figure*}[t]
	\begin{adjustbox}{addcode={\begin{minipage}{\width}}{\caption{\label{fig:multispec2} 
		In-beam $\gamma$-ray spectra from the 55$^\circ$ detector for the $E_p$ = 857-1401\,keV runs. See caption of \fig{fig:multispec1}. The 6792\,keV secondary $\gamma$-rays at $E_p$ = 1115\,keV and 1300\,keV are discussed in the text.
		}\end{minipage}},rotate=90,center}
	\includegraphics[angle = 0,width=\textheight]{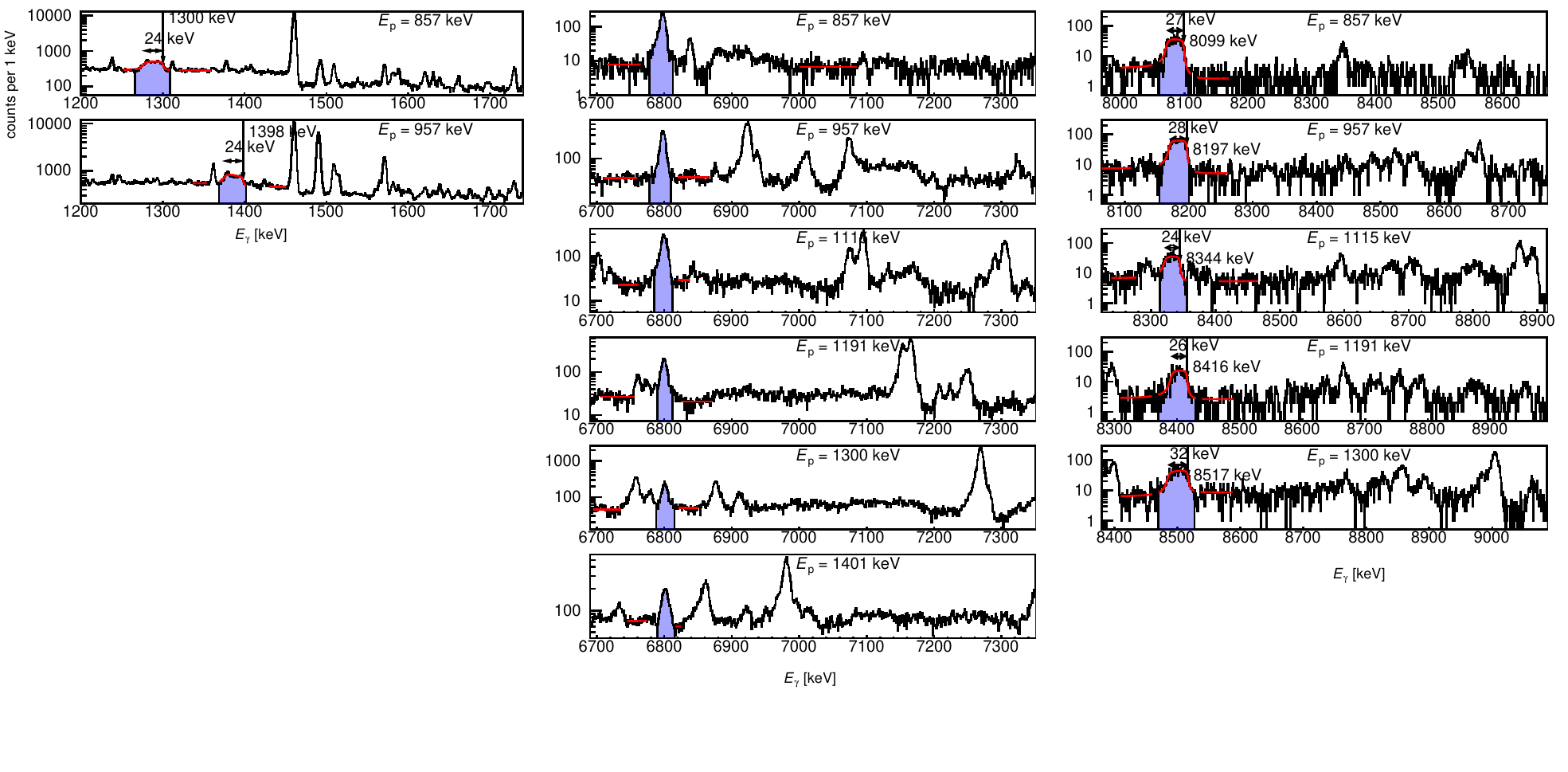}
	\end{adjustbox}
\end{figure*}

\begin{itemize}
\item At $E_p$ = 533\,keV, the ground state primary is affected by background due to the 23\,keV wide $^{13}\mathrm{C(p,}\gamma)^{14}\mathrm{N}$ resonance at $E_{\rm p}$ = 551\,keV \cite{Ajzenberg13_15_91-NPA}. In order to treat this background, the shape of the detector response has been simulated by GEANT4 and subtracted from the observed spectrum (\fig{fig:spec533}). After subtraction of the Compton edge based on the simulation (60\% and 30\% of the raw counts for the 55$^\circ$ and 90$^\circ$ detector, respectively), the ground state capture peak clearly emerges, albeit on top of a remaining continuum. The position and width of the peak coincide with what is expected from the resonance scan (\Sec{subsec:nrra}). The error bar for these data points is conservatively increased by 30\% of the subtracted counts.
\item At $E_p$ = 1115 keV, there is a $\gamma$-ray exactly 511\,keV above the 6792\,keV secondary, so that its single-escape peak had to be subtracted based on the known single-escape/full-energy peak ratio, giving 16\% correction.
\item At $E_p$ = 1301 keV, a $\gamma$-ray at 9010\,keV lies 511\,keV above the ground state primary. This peak is tentatively assigned to the $^{18}$O(p,$\gamma$)$^{19}$F reaction. Its single-escape peak had to be subtracted based on the known single-escape to full-energy peak ratio, giving 50\% correction.
\end{itemize}

For the ground state peak, the yield has been corrected down by 0.4-1.5\% for the summing-in effect. This correction has been estimated based on the $\gamma$-ray detection efficiency and the ratio between the astrophysical S-factors for the 6.79 and ground state transitions, and a conservative 20\% relative uncertainty was assumed for the correction. Based on the total $\gamma$-ray efficiency, the summing-out correction for the 6792\,keV secondary peak was found to be even lower, always below 0.6\%. 

\clearpage
\FloatBarrier{}
\subsection{\label{subsec:angdist} Angular information gained}

\begin{figure}[t]
\includegraphics[angle = 0, width=\columnwidth]{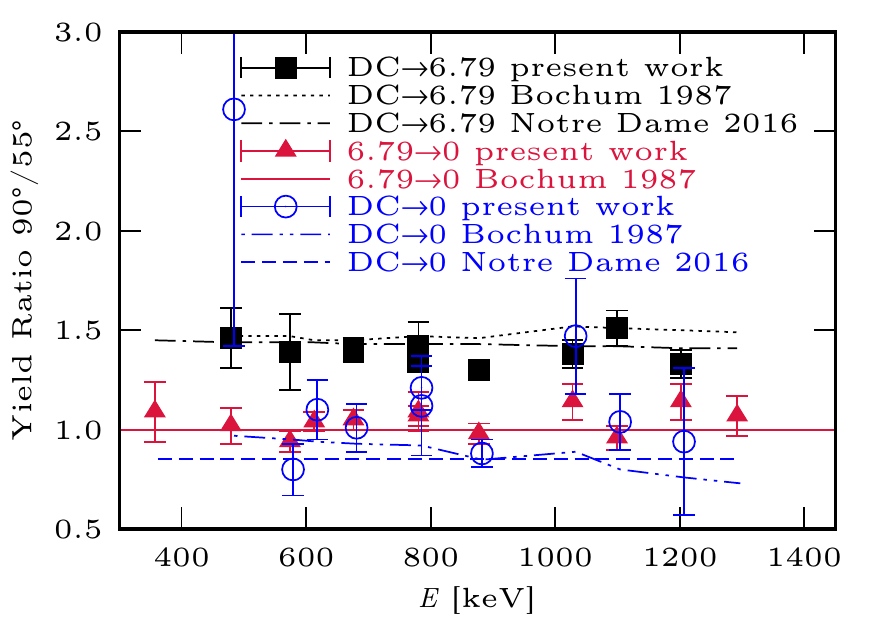}
\caption{\label{fig:Angratio} 
Ratio of the efficiency-corrected $\gamma$-ray yields from the 90$^\circ$ and 55$^\circ$ detectors from the present work, compared with previous results from Bochum \cite{Schroeder87-NPA} and Notre Dame \cite{Li16-PRC}. For the latter case, the smoothly varying external-capture calculations based on their R-matrix fit are shown \cite{Li16-PRC}.
}
\end{figure}

The angular distribution of the off-resonant $^{14}\textrm{N(p,}\gamma\textrm{)}^{15}\textrm{O}$ $\gamma$-ray yield was measured in a wide energy range in the Bochum experiment and presented in the form of Legendre coefficients \cite{Schroeder87-NPA}. The distribution was recently re-measured at Notre Dame \cite{Li16-PRC}. The angular data from these two works are different for the primary $\gamma$-rays. For the $\gamma$-rays from direct capture to the ground state and to the 6792\,keV excited state, Notre Dame reported a non-negligible coefficient for the $P_{1}$ Legendre polynomial, which lead to up to 40\% forward-backward asymmetry \cite{Li16-PRC}. Bochum had reported such a forward-backward asymmetry for ground state capture, but not for capture to  the 6792\,keV excited state \cite{Schroeder87-NPA}.

The present data is not very sensitive to the angular distribution. The first-order Legendre polynomial vanishes at 90$^\circ$, $P_1(\cos 90^\circ)$~=~0. The second-order Legendre polynomial vanishes at 55$^\circ$, $P_2(\cos 55^\circ)$~=~0. The secondary $\gamma$-ray of the 6.79~MeV transition was reported to be isotropic by Bochum \cite{Schroeder87-NPA} and not studied by Notre Dame \cite{Li16-PRC}. The efficiency-corrected ratio of the yields of the two detectors is consistent with isotropy (\fig{fig:Angratio}). 

For ground state capture, the different reported Legendre coefficients from Bochum and Notre Dame partially cancel out in the yield ratio, and the present data are in fair agreement with the yield ratio expected based on these works (\fig{fig:Angratio}). The only outlier is the ground state data point at $E_p$ = 1115\,keV, just above the $E$ = 987\,keV resonance. The 90$^\circ$ yield is 50\% higher than the 55$^\circ$ yield, whereas the angular distributions by both Bochum and Notre Dame predict it to be lower. It is noted that  it was found previously that the Legendre coefficients from the experimental data and also from the R-matrix fit vary strongly with energy near this resonance \cite{Li16-PRC}.

For the data analysis, no angular correction is made for the 6.79~MeV secondary $\gamma$-ray. For the primary $\gamma$-ray to the 6.79~MeV level, the data are corrected with the measured angular coefficients by the recent Notre Dame experiment \cite{Li16-PRC}. For the ground state primary $\gamma$-ray, the previous coefficients from the Bochum experiment \cite{Schroeder87-NPA} are used instead. They are consistent with Notre Dame \cite{Li16-PRC} and more easily accessible in the paper. 

\subsection{\label{subsec:sfactor}Determination of the cross section and astrophysical S-factor}

The experimentally observed yield $Y(E_p)$ and the sought after cross section are connected by the following relation:
\begin{equation}\label{eq:yield}
Y(E_p) = \int_{E_p^{\rm corr}}^{E_p^{\rm corr} - \Delta E_p} \frac{\sigma(E_{\rm lab})}{\epsilon_{\rm eff}^{14}(E_{\rm lab)}} dE_{\rm lab}
\end{equation}
which was numerically integrated assuming the plateau Ti:N stoichiometry from the ERD analysis (\Sec{subsec:erda}). The proton beam energy $E_p^{\rm corr}=E_p-\Delta E_p^{\rm C}$ in \eq{eq:yield} was obtained by subtracting $\Delta E_p^{\rm C}$ = 2-10 keV energy loss in the initial, 3-70$\times$10$^{16}$~at/cm$^2$ thick carbon layer (\Sec{subsec:nrra}) from the initial proton beam energy. This reduction in $E_p$ lead to an increase of 0.2-2.5\% in the astrophysical S-factor. Conservatively a 30\% relative uncertainty was assumed for the correction, leading to up to 0.8\% error resulting from this effect.

As a cross check, the analysis was repeated by again numerically integrating Eq.~(\ref{eq:yield}) but starting directly from $E_p$ and taking into account the carbon layer, and all other minor impurities detected, based on their depth-dependent concentrations from the ERD analysis (\Sec{subsec:erda}). The difference in results with the standard analysis methods was lower than the statistical uncertainty. 

The resulting S-factor depends by necessity on the assumed shape of the S-factor curve. The calculation was repeated first with the Solar Fusion II (SFII) S-factor curve, then assuming a flat S-factor, showing differences of $<$1\% in the final S-factor, and this effect was thus neglected. The effective energy assigned to the S-factor was taken as the median energy \cite{Rolfs88-Book} of the integrand of \eq{eq:yield}. The final astrophysical S-factor values are summarized in Table~\ref{tab:s-factor}.

\begin{table}[t]
\caption{\label{tab:s-factor} $^{14}\textrm{N(p,}\gamma\textrm{)}^{15}\textrm{O}$ S-factors for capture to the 6.79~MeV excited state and for capture to the ground state in $^{15}$O, as a function of the effective center of mass energy $E$.}
\begin{ruledtabular}
\begin{tabular}{r*{4}{c}}
\multicolumn{1}{c}{$E$}	&	$S_{6.79}(E)$	&	$\Delta S^\textrm{stat}_{6.79}$	&	$S_\textrm{gs}(E)$	&	$\Delta S^\textrm{stat}_\textrm{gs}$	\\
 $[$keV]	& [keV barn]	& [keV barn]	& [keV barn]	& [keV barn]	\\
\hline
357		&	1.27	&	0.10	&	-	&	-	\\
479		&	1.12	&	0.07	&	0.19	&	0.07	\\
574		&	1.17	&	0.04	&	0.26	&	0.02	\\
613		&	1.05	&	0.04	&	0.22	&	0.02	\\
676		&	1.14	&	0.04	&	0.24	&	0.02	\\
780		&	1.00	&	0.04	&	0.26	&	0.03	\\
780		&	1.07	&	0.03	&	0.31	&	0.03	\\
877		&	1.06	&	0.03	&	0.42	&	0.02	\\
1028	&	1.17	&	0.05	&	0.30	&	0.05	\\
1099	&	1.23	&	0.04	&	0.29	&	0.03	\\
1202	&	1.21	&	0.07	&	0.23	&	0.05	\\
1292	&	1.15	&	0.07	&	-	&	-	\\
\end{tabular}
\end{ruledtabular}
\end{table}

\subsection{\label{subsec:uncertainties}Uncertainties}

The uncertainties of the present data points (Table \ref{tab:uncertainties}) are divided in two groups: The first group (systematic uncertainties) are scale factors that would, at least in principle, affect all the present data points uniformly. The second group (statistical uncertainties) of uncertainties affect each data point randomly and may thus have a different sign for each data point. Only the latter uncertainties should be used when e.g. gauging the appropriateness of an R-matrix fit. The former uncertainties will then determine the additional scaling uncertainty of the fit result.

\begin{figure*}[bt]
\includegraphics[angle = 0,width=0.9\textwidth]{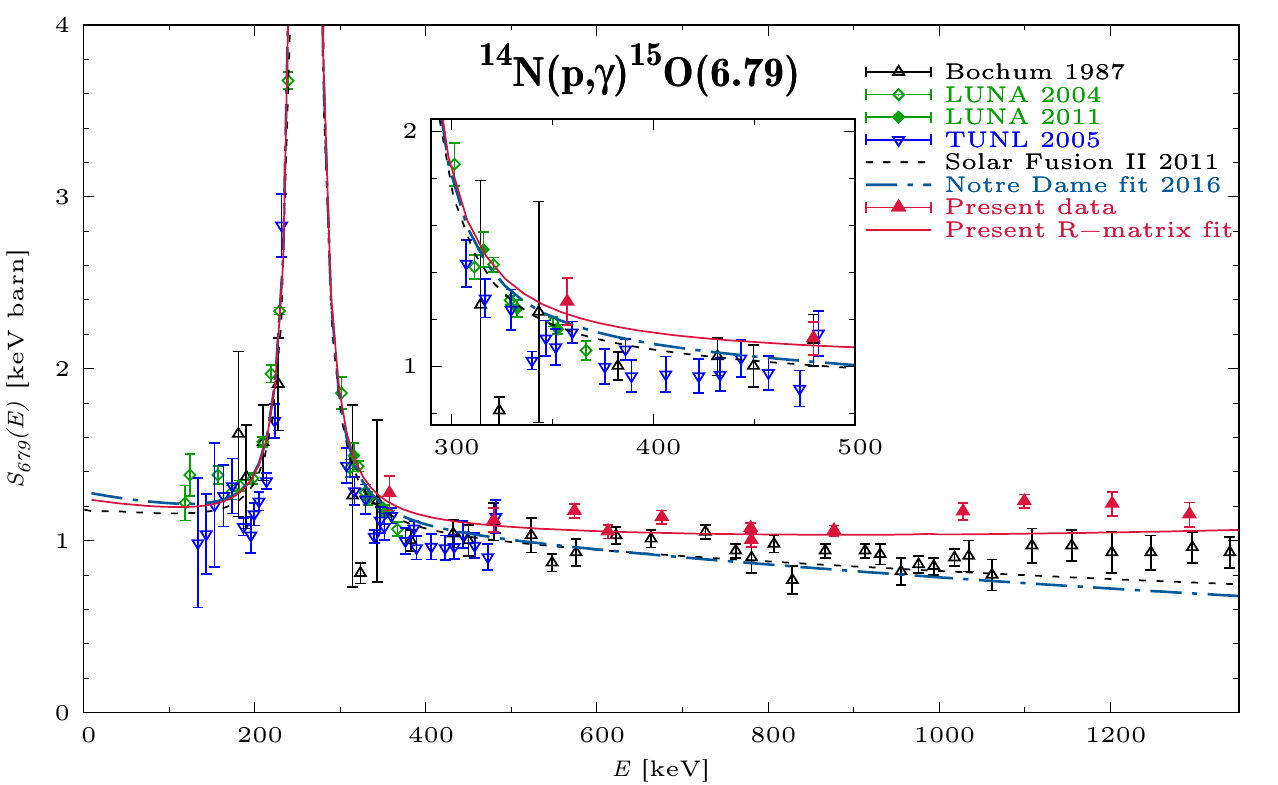}
\caption{\label{fig:sf679} Astrophysical S-factor for the 6.79~MeV transition in $^{14}\textrm{N(p,}\gamma\textrm{)}^{15}\textrm{O}$ from the literature \cite{Schroeder87-NPA,Formicola04-PLB,Imbriani05-EPJA,Marta08-PRC,Marta11-PRC,Runkle05-PRL} and from the present work. The data from Refs.~\cite{Schroeder87-NPA,Formicola04-PLB,Imbriani05-EPJA,Runkle05-PRL} have been renormalized as in Ref.~\cite{Adelberger11-RMP} for the 13.1~meV strength of the 259~keV resonance. The R-matrix fits by SFII \cite{Adelberger11-RMP}, Notre Dame \cite{Li16-PRC}, and from the present work (\Sec{sec:rmatrix}) are also shown.}
\end{figure*}

The largest systematic uncertainty, 6\%, stems from the determination of the target composition, here expressed as the effective stopping power, by the ERD method (\Sec{subsec:erda}). This determination is made separately for each target studied here, however using one and the same beam, detectors, and analysis method. Therefore it is conservatively assumed that the ERD uncertainty has a scale factor component of 6\% common to all data points (due to the calibration of the ERD apparatus used) and a statistical component that is target dependent of 5\%. The $\gamma$-ray detection efficiency contributes 3\% error (\Sec{subsec:gammadetectors}). The 5\% uncertainty due to the angular corrections \Sec{subsec:angdist} is estimated based in the analysis of the yield ratios (\fig{fig:Angratio}). The beam intensity is estimated to be known to 1\%, due to remaining imperfections of the Faraday cup used. The beam energy calibration affects the conversion of the measured yield to the astrophysical S-factor. It contributes negligibly to the error budget, always less than 0.3\%, and is therefore not listed in the table. Similarly, the error due to summing corrections (\Sec{subsec:gammaspectra}) was always 0.3\% or below and therefore left out of the table.

The energetic target thickness is determined from the resonance scans of the target, and its uncertainty is mainly given by the statistical error from the fit curve obtained, meaning it has to be treated like a statistical uncertainty contributing up to 3\% error. The main statistical uncertainty is from the $\gamma$-ray counting statistics and from the background subtraction, where applicable. For the 6.79~MeV transition, the statistical uncertainty is between 2\% and 11\%. For the ground state transition, due to the lower statistics, higher uncertainties of typically 5-18\% are found. 

There are two cases with higher statistical uncertainty in the ground state transition: At $E=479$~keV ($E_p$~=~533~keV) where the ion beam induced background subtraction plays a major role, for the ground state primary $\gamma$-ray a statistical error bar of 37\% (28\%) is found in the 55$^\circ$ (90$^\circ$) detector, leading to a total statistical uncertainty of 36\% for the weighted average of the two, taking the uncertainties in the angular correction into account. The second case is $E$~=~1202~keV with the subtraction of a single-escape line leading to a statistical uncertainty of 20\% in the weighted average.

\begin{table}[tb]
\caption{\label{tab:uncertainties} Error budget for the astrophysical S-factor, in percent. See text for the $E=479$ and 1202~keV data points and for further details. }
\begin{ruledtabular}
\begin{tabular}{lD{.}{.}{1}l}
Uncertainty 			& \multicolumn{1}{c}{syst.}	& \multicolumn{1}{c}{stat.} \\
\hline
Effective stopping power (\Sec{subsec:erda})					& 6	& 5 \\
$\gamma$-ray detection efficiency (\Sec{subsec:gammadetectors})	& 3	&  \\
angular correction (\Sec{subsec:angdist})					& 5 & \\
Beam intensity (\Sec{subsec:chamber})						& 1	&  \\
Effective beam energy $E$						&	& 0.2-1.2 \\
Energetic target thickness $\Delta E_p$						&	& 3 \\
Count rate, ground state 								&	& 5-14 \\
Count rate, 6.79~MeV									&	& 2-11 \\
\hline
Total, ground state										& 8	& 5-15 \\ 
Total, 6.79~MeV										& 8	& 3-8 \\ 
\end{tabular}
\end{ruledtabular}
\end{table}

\subsection{\label{subsec:Discussion}Discussion of the results}

The present data for capture to the 6.79~MeV state cover the energy range $E$ = 357-1292~keV (\fig{fig:sf679}). In the S-factor representation, they display a linear behavior over the entire energy range studied, with the exception of a slight increase caused by the high-energy tail of the strong 259~keV resonance. 

In the low-energy region between 300-500~keV, where many precise data points are available from the LUNA \cite{Schroeder87-NPA,Formicola04-PLB,Imbriani05-EPJA,Marta08-PRC,Marta11-PRC} and TUNL \cite{Runkle05-PRL} groups, the two lowest-energy data points agree, within their statistical error bars, with these previous data. They are also in agreement with the R-matrix fits from SFII and Notre Dame \cite{Adelberger11-RMP,Li16-PRC} and with the Bochum data \cite{Schroeder87-NPA}. 

The slope of the present data to higher energies, however, is different, with a slight rise that is not seen in the previous Bochum data \cite{Schroeder87-NPA}. In the $E$=1000-1300 keV range, the present data are on average 25\% higher than Bochum. There are no cross section data available from the Notre Dame experiment in this energy range \cite{Li16-PRC}.

For capture to the ground state in $^{15}$O, there are fewer data points in the present work, due to the smaller absolute size of the cross section for this transition and due to the $^{13}$C(p,$\gamma$)$^{14}$N background. Therefore, there is only one point of overlap with the low-energy data, not two as for the case of the 6.79~MeV transition. 

\begin{figure}[tb]
\includegraphics[angle = 0,width=\columnwidth]{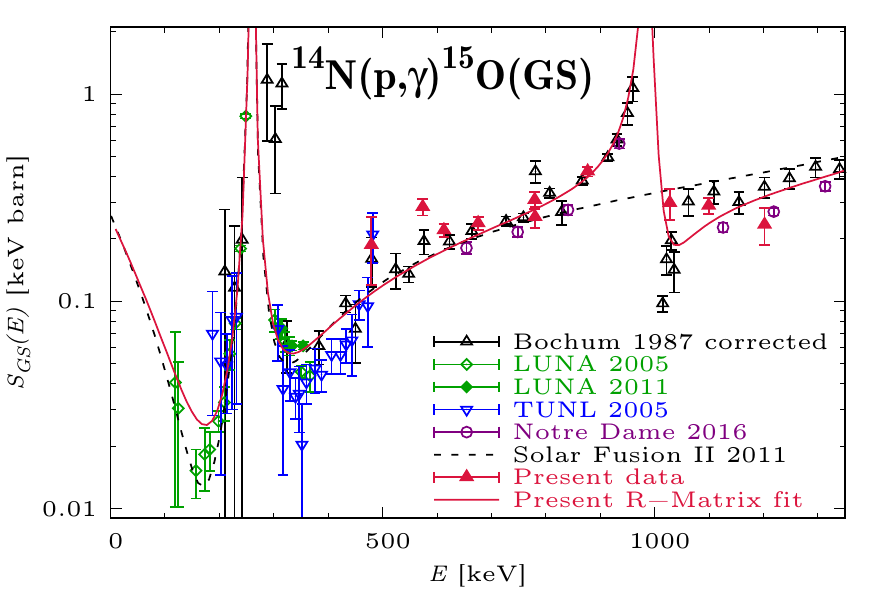}
\caption{\label{fig:sfgs} Astrophysical S-factor for the ground state transition in $^{14}\textrm{N(p,}\gamma\textrm{)}^{15}\textrm{O}$ from the literature \cite{Schroeder87-NPA,Formicola04-PLB,Imbriani05-EPJA,Marta08-PRC,Marta11-PRC,Runkle05-PRL,Li16-PRC} and from the present work. Normalization as in Fig.~\ref{fig:sf679}. The Bochum data \cite{Schroeder87-NPA} are shown corrected \cite{Formicola04-PLB,Adelberger11-RMP} for the summing-in effect. The R-matrix fits by SFII \cite{Adelberger11-RMP} and from the present work are also shown.}
\end{figure}

The present ground state data are generally not far from the scale and slope of the corrected and renormalized Bochum data (Fig.~\ref{fig:sfgs}). There are three exceptions, the first being the data point at $E$~=~574~keV ($E_p$~=~640~keV) which sits on top of a relatively flat Compton region caused by the $^{13}$C(p,$\gamma$)$^{14}$N background peak (see \fig{fig:multispec1}). 

Second, the data point at $E$ = 1028\,keV just above the $E$ = 987\,keV resonance may have a problem with the angular correction as discussed in \Sec{subsec:angdist}. In order to take this problem into account, the error bar is enlarged to cover also the value found when only analyzing the 55$^\circ$ detector. There is no obvious explanation for the remaining distance to the Bochum data, and to the R-matrix fit. There is no Notre Dame data point at this energy. 

The third exception is the highest-energy data point, which is corrected down by 50\% for a single-escape peak (see above section~\ref{subsec:gammaspectra}). This data point is significantly lower than Bochum and than the R-matrix curve, but close to a Notre Dama data point at similar energy.

\section{\label{sec:rmatrix}R-matrix fit}

In order to estimate their impact on the low-energy extrapolated S-factor, the present data were then included in a limited R-matrix fit, using the AZURE2 code \cite{Azuma10-PRC}. Different from other recent work \cite{Azuma10-PRC,Adelberger11-RMP,Li16-PRC}, no full R-matrix fit is attempted here. 
In particular, no elastic scattering data \cite{deBoer15-PRC} are included, and angular distribution data are only used to correct the absolute cross section, not for the R-matrix fit.

The present fit 
therefore serves mainly as an illustration of the possible effects of the present new data on the extrapolated cross section at low energy. The fit is limited to those quantities that are expected to have an effect on either the normalization or the slope of the non-resonant S-factor curve, or on both: The asymptotic normalization coefficient (ANC) of the 6.79~MeV level and the widths of the so-called background poles.

The selection, and renormalization, of cross section data included in the fit routine follows SFII \cite{Adelberger11-RMP}: LUNA 2004-2005 \cite{Formicola04-PLB,Imbriani05-EPJA}, scaled by 1.02. -- LUNA 2008-2011 \cite{Marta08-PRC,Marta11-PRC}, no rescaling. -- TUNL \cite{Runkle05-PRL}, scaled by 0.97. The present data are included without normalization, as they do not depend on the strength of the 259~keV resonance. For all data sets, S-factor values close to sharp resonances were excluded by the same criterion as in SFII \cite{Adelberger11-RMP}, in order to limit the impact of data points where the generally low energy uncertainty may lead to significant deviations from the fit curve. The corrected Bochum \cite{Schroeder87-NPA} data were used in SFII (scaled by 0.92) but not here, instead they are replaced by the present new data. 

It has been shown previously \cite{Mountford14-NIMA} that the AZURE2 code used here gives similar results to the hitherto used \cite{Angulo01-NPA,Formicola04-PLB,Imbriani05-EPJA,Marta08-PRC,Marta11-PRC,Adelberger11-RMP} Descouvemont code \cite{Descouvemont10-RPP}. 

The parameters that are kept fixed in the present simplified fit are discussed below.

First, the very low channel radius of 4.2\,fm used previously to fit  $^{14}\textrm{N(p,p)}^{14}\textrm{N}$ scattering data \cite{deBoer15-PRC} was tested here, but it led to an imaginary number for the external partial width of the background pole, so this attempt was discarded. A channel radius of 5.5\,fm was then used here (the same number as in SFII \cite{Adelberger11-RMP} and in \Ref{Li16-PRC}), which led to a real number for the background pole width and improved $\chi^2$ by 10\,\%. 

Second, the asymptotic normalization coefficients for ground state capture by Mukhamedzhanov \cite{Mukhamedzhanov03-PRC} are used here, converted to the AZURE2 coupling scheme \cite{Azuma10-PRC}: $C_{p,1/2}$~=~(0.23$\pm$0.01)~fm$^{-1/2}$ and $C_{p,3/2}$~=~(7.3$\pm$0.4)~fm$^{-1/2}$. For the latter value, SFII \cite{Adelberger11-RMP} and Notre Dame \cite{Li16-PRC} use a slightly higher value of (7.4$\pm$0.5)~fm$^{-1/2}$.

Third, the level energies are taken from the Ajzenberg-Selove evaluation \cite{Ajzenberg13_15_91-NPA}, except where updated by LUNA \cite{Imbriani05-EPJA}, and kept fixed in the fit. This is different from Notre Dame, where Ajzenberg-Selove energies are used, with the level energies above 7.56\,MeV being allowed to vary in the fit \cite{Li16-PRC}.

Fourth, in order to preserve the information contained in the S-factor values close to sharp resonances (which are excluded from the fit for the reasons given above), some parameters of the strong 259~keV resonance are kept fixed at the \Ref{Azuma10-PRC} values, namely, the proton width and partial, internal $\gamma$-ray widths for decay to the ground and 6.79~MeV excited states: $\Gamma_p^{259}$ = 1.0~keV, $\Gamma_{\gamma,0}^{259}$~=~0.4~meV, $\Gamma_{\gamma,6.79}^{259}$~=~9.6~meV. These values are close to Notre Dame values, except for $\Gamma_{\gamma,0}^{259}$~=~0.65~meV \cite{Li16-PRC}, a difference which however has only limited impact at $E$~$<$~200~keV.

Fifth, for the resonances at 0.987 and 2.187~MeV, starting values from Notre Dame were used \cite{Li16-PRC} only marginally varied. 

For easier reference, all the R-matrix parameters that have been changed with respect to  Ref. \cite{Li16-PRC} are listed in \tab{tab:fitparam}.

\begin{table}[tb]
\caption{\label{tab:fitparam} Summary of parameters used in the R-matrix fit. See text for details.}
\begin{ruledtabular}
\begin{tabular}{lrr}
		& Present work	& Ref. \cite{Li16-PRC}\\
	\hline
	R							& 5.5~fm			& 5.5~fm\\
	\hline
	C$_{p,1/2}$					& 0.23~fm$^{-1/2}$	& 0.23~fm$^{-1/2}$\\
	C$_{p,3/2}$					& 7.3~fm$^{-1/2}$	& 7.4~fm$^{-1/2}$\\
	\hline
	$E_{x,1/2+}$				& 5180.8~keV		& 5183.0~keV\\
	$E_{x,3/2-}$				& 6172.3~keV		& 6176.3~keV\\
	$E_{x,3/2+}$				& 6791.7~keV		& 6793.1~keV\\
	\hline
	$\Gamma_{p}(259)$			& 1.0~keV			& 0.96~keV\\
	$\Gamma_{\gamma,0}(259)$	& 0.4~meV			& 0.65~meV\\
	$\Gamma_{\gamma,6792}(259)$	& 9.6~meV			& 9.3~meV\\
	\hline
	$E_{x,3/2+}$			& 8289~keV			& 8285~keV\\
	$\Gamma_{p}(2187)$			& 71~keV			& 89~keV\\
	$\Gamma_{\gamma,0}(\rm BGP, 3/2^{+})$		&492~eV	&220~eV\\
	$\Gamma_{\gamma.0}(\rm BGP, 5/2^{-})$		&675~eV	&-\\
\end{tabular}
\end{ruledtabular}
\end{table}
For capture to the 6.79~MeV excited state, the resulting fit (red solid curve in \fig{fig:sf679}) shows a somewhat different slope in the $E$~=~400-1300~keV range than SFII and Notre Dame, which are both below the present experimental data in this energy range \cite{Adelberger11-RMP,Li16-PRC}. This effect is most visible  at $E$~=~1000-1300~keV. 

Despite these non-negligible differences at high energies, the picture is more consistent at low, astrophysical energies. There, the present fit comes out only about 2\% higher than SFII. It should be kept in mind that Notre Dame reported a relatively high zero-energy S-factor for this transition, $S_{6.79}(0)$ = 1.29$\pm$0.04(stat)$\pm$0.09(syst)~keV~barn \cite{Li16-PRC}, higher than but still consistent with the SFII value of 1.18~keV~barn. The present result of $S_{6.79}(0)$~=~1.24$\pm$0.11~keV~barn lies between SFII and Notre Dame and is consistent with both. The systematic uncertainty of $S_{6.79}(0)$ derives from the 9\% systematic (scale) uncertainty of the present data points. The statistical uncertainty of $S_{6.79}(0)$ has been studied by repeating the R-matrix fit with a grid of different values for the two background poles and the ANC and was found to be 0.02~keV~barn, negligible when compared to the systematic uncertainty. The total uncertainty of $S_{6.79}(0)$ is thus $\pm$0.11~keV~barn.

For capture to the ground state, the present fit results in a zero-energy extrapolation of $S_{\rm GS}(0)$ = 0.19$\pm$0.05~keV~barn, lower than the SFII value of 0.27$\pm$0.05~keV~barn. The higher upper limit of the error band recently suggested by Notre Dame with its value of 0.42$\pm$0.04(stat)$^{+0.09}_{-0.19}$(syst) is not confirmed here. Interestingly, in the depression at $E$~$\sim$~300~keV, the present fit seems to result in a compromise between the virtually summing-free LUNA data taken with a segmented detector \cite{Marta08-PRC,Marta11-PRC} and the TUNL and remaining LUNA data with their summing issues \cite{Formicola04-PLB,Imbriani05-EPJA,Runkle05-PRL}. SFII shows the same behaviour in this energy range but is lower below the 259\,keV resonance where only a few data from LUNA exists. This may explain the lower zero-energy extrapolation in the present work when compared to SFII.  

For the uncertainty of  $S_{\rm GS}(0)$, the scaling uncertainty of the present data points makes only a negligible contribution, mainly due to the much stronger influence of the low-energy data points from LUNA and TUNL. The statistical uncertainty dominates. When repeating the R-matrix fit with a grid of different values for the subthreshold resonance strength and the background pole, a statistical uncertainty of $\pm$0.01~keV~barn is found.

\section{\label{sec:summary}Summary and outlook}

A new measurement of the cross section of the $^{14}$N(p,$\gamma$)$^{15}$O reaction was undertaken based on the analysis of two transitions. S-factor data were obtained by in-beam $\gamma$-ray spectroscopy at twelve energies between 357-1292~keV for capture to the 6.79~MeV excited state in $^{15}$O and at ten energies between 479-1202~keV for capture to the ground state in $^{15}$O. The absolute cross section was determined, normalized to a target composition obtained by the elastic recoil detection technique.

The new data are not far from the previous wide energy range excitation function by the Bochum group \cite{Schroeder87-NPA}, which had recently been questioned due to correction and renormalization issues. However, for the strongest transition, capture to the 6.79~MeV excited state, the present data show a somewhat higher slope than Bochum towards the higher-energy end. 

The impact of the new data on low astrophysical energies is gauged by a preliminary R-matrix fit. 

For the 6.79~MeV transition, the resulting zero-energy extrapolated S-factor, $S_{6.79}(0)$ = 1.24$\pm$0.11(syst)$\pm$0.02(stat)~keV~barn, lies between the recently reported Notre Dame \cite{Li16-PRC} and the previously accepted SFII \cite{Adelberger11-RMP} extrapolated values. It seems that the low energy extrapolation is robust even when taking the somewhat higher, present high-energy 6.79~MeV data into account.

For the ground state transition, the present extrapolated value of $S_{\rm GS}(0)$ = 0.19$\pm$0.05(syst)$\pm$0.01(stat)~keV~barn is lower than but still consistent with the Notre Dame and SFII values. 

Summarizing, the 6.79~MeV transition may be excluded as a source of significant uncertainty for the total extrapolated cross section.
However, the situation is different for the weaker transitions, including but not limited to ground state capture. For these cases, a new comprehensive data set connecting the precise low-energy LUNA \cite{Formicola04-PLB,Imbriani05-EPJA,Marta08-PRC,Marta11-PRC} with the wide energy range Bochum data points is still missing. Due to the long running times and low counting rates, such data can best be provided at one of the upcoming higher-energy underground accelerators \cite{Guglielmetti14-DARK,Robertson16-EPJWOC,Liu16-EPJWOC,Bemmerer16-arXiv}.

\begin{acknowledgments}

The authors are indebted to Andreas Hartmann (HZDR) for technical support, to Mario Steinert (HZDR) for the production of TiN targets, and to Dr Richard J. deBoer (University of Notre Dame) for his kind assistance with the AZURE2 code. ---
This work was supported in part by the Helmholtz Association (grant number VH-VI-417, Nuclear Astrophysics Virtual Institute NAVI), the Helmholtz Detector Systems and Technologies platform, Deutsche Forschungsgemeinschaft (DFG, grant number BE 4100/2-1), and by the European Union (FP7-SPIRIT, Contract No. 227012). L.W. and M.T. gratefully acknowledge fellowships by the TU Dresden Graduate Academy. L.W. is grateful for support to attend an R-matrix workshop by the JINA Center for the Evolution of the Elements (National Science Foundation, Grant No. PHY-1430152).

\end{acknowledgments}


\end{document}